\documentstyle[prd,aps,multicol,tighten,epsf,rotate]{revtex}

\begin{document}
\draft

\preprint{\vbox{\it 
                        \null\hfill\rm    IP-BBSR/97-27, June'97}\\\\}
%
\title{Simulation of Vortex-Antivortex Pair Production in a Phase 
Transition with Explicit Symmetry Breaking} 
\author{Sanatan Digal, Supratim Sengupta and Ajit M. 
Srivastava \footnote{E-mail :\\digal@iopb.stpbh.soft.net\\ 
supratim@iopb.stpbh.soft.net\\ ajit@iopb.stpbh.soft.net}}
\address{\it Institute of Physics\\
Sachivalaya Marg, Bhubaneswar--751005, INDIA}
\maketitle
\widetext
\parshape=1 0.75in 5.5in
\begin{abstract}
 We carry out numerical simulation of the formation of U(1) global 
vortices in a first order phase transition in 2+1 dimensions in the 
presence of small explicit symmetry breaking. Bubbles of broken symmetry 
phase are randomly nucleated, which grow and coalesce. Vortices 
form at junctions of bubbles via standard Kibble mechanism as well as 
due to a new mechanism, recently proposed by us, where defect-antidefect 
pairs are produced due to field oscillations. In a simulation 
involving nucleation of 63 bubbles, with bias in phase distribution 
inside bubbles arising from explicit symmetry breaking, we find that not 
a single vortex/antivortex is produced via the Kibble mechanism, while 
the new mechanism leads to production of 104 vortices and antivortices.
Even without biasing the phase distribution inside bubbles, the 
vortex production is completely dominated by this new mechanism,
which accounts for the production of about 80\% of the vortices and 
antivortices, remaining 20\% being produced via the Kibble mechanism.  
We study the dependence of the effectiveness of the new mechanism on 
the magnitude of explicit symmetry breaking, as well as on the 
nucleation rate of bubbles. We also study the effect of damping on 
this mechanism and show that damping suppresses this mode of vortex 
production. 
\end{abstract}
\vskip 0.125 in
\parshape=1 -.75in 5.5in
\pacs{PACS numbers: 98.80.Cq, 61.30.Jf, 12.39.Dc}
\begin{multicols}{2}
\narrowtext

\vskip .2in
\centerline {\bf I. INTRODUCTION}
\vskip .1in

 There has been a renewed interest recently in studying the formation 
of topological defects in phase transitions. It is clearly important 
to understand various processes which could be responsible for 
defect formation as then only one can hope to get a true understanding 
of the distribution of defects and its time evolution. These issues are 
of great importance not only for condensed matter physics but also in 
the context of particle physics models of the early Universe where 
topological defects are supposed to play an important role in the 
evolution of the Universe and in the structure formation \cite{shlrd}.
 
In conventional studies only two distinct types of processes have been 
considered to be important in studying the formation of topological 
defects in phase transitions. One of these is based on thermal production 
of defects which leads to defect density which is suppressed by the 
Boltzmann factor \cite{thrm}. The other process is based on the formation 
of a kind of domain structure during phase transition with defects forming 
at the junctions of these domains, and is typically known as the Kibble 
mechanism \cite{kbl}. Kibble mechanism was originally proposed for 
studying defect formation in the early Universe. However, the mechanism 
as such has complete general applicability and in fact has been recently 
verified by studying defect formation in certain condensed matter 
systems \cite{zurk,lcrs}.   [Though, there are nontrivial issues for 
the case when gauge fields are also present. See ref. \onlinecite{cs1} 
and references therein.]

 An important aspect of the Kibble mechanism is that it does not crucially
depend on the dynamical details of the phase transition. For example, the
number of defects (per domain) produced via the Kibble mechanism 
depends only on the topology of the order parameter space and 
on spatial dimensions. Dynamics plays a role here only in determining the
relevant correlation length, which in turn determines the domain size,
affecting net number of defects produced in a given region. Still, the
number of defects per domain is entirely independent of the dynamics.  
[Apart from some special situations, e.g. in a very slow first order
transition, see \onlinecite{brl}.]

 Recently, a new mechanism for defect production has been proposed by 
us \cite{dgl1,dgl2,dgl3} where the dynamics of the order parameter field 
plays a very important role. Here, defect-antidefect pairs are
produced due to strong oscillations of the field. We showed that whenever 
the field passes through zero magnitude, while oscillating, (in a region
where the field is non-uniform), a defect-antidefect pair gets created.
This essentially explains why the dynamics plays a major role in this 
mechanism since the nature and the strength  of field oscillations during 
a phase transition will in general depend on various dynamical details, 
such as the order of the transition, time duration of the quench etc. We 
find that the number of defects (say per domain) produced via this 
mechanism can vary strongly as different parameters of the transition 
are changed. 

 This mechanism was first proposed by two of us \cite{dgl1} for systems 
in the presence of small explicit symmetry breaking. Using numerical 
simulation of two bubble collisions in 2+1 dimensions for the case of 
U(1) global vortices, it was shown in ref. \onlinecite{dgl1} that a 
very large number of vortex-antivortex pairs can be created via this 
mechanism due to field oscillations enhanced by the explicit symmetry
breaking. For example, in one case it was found that ten, well 
separated,  vortices and antivortices were created in a single two 
bubble collision. It was later shown by us in ref. \onlinecite{dgl2}, 
that this mechanism was not restricted only to systems with explicit 
symmetry breaking, and is in fact very  general. It applies even when 
there is no explicit symmetry breaking and is applicable to all sorts 
of topological defects. We also argued that this mechanism should be 
operative even in a second order phase transition involving quenching 
from a sufficiently high temperature. However, as this mechanism 
depends sensitively on dynamics, we found that defect production for the
case of zero explicit symmetry breaking was not very prominent compared
to the case with explicit symmetry breaking studied in 
ref. \onlinecite{dgl1}.  [At least for the range of bubble separations 
considered in ref. \onlinecite{dgl1,dgl2}.] 

  Here we mention the work of Copeland and Saffin \cite{cs1,cs2} who have
considered the production of strings in gauge theories. They showed in
ref. \onlinecite{cs1}, for Abelian Higgs model, that the geodesic rule 
may get violated in the collision of two bubbles and that vortex-antivortex 
pairs can form in the region of coalescence. In their case also the 
oscillations of field played crucial role, though due to the presence of 
gauge field the dynamics of the phase $\theta$ had extra features. For 
example, the presence of gauge fields provides a driving force for 
$\theta$. The dynamics of vortex production in ref. \onlinecite{cs1} is, 
in this sense, similar to the case of explicit symmetry breaking 
discussed in ref. \onlinecite{dgl1}.

 The studies in ref. \onlinecite{dgl1,dgl2} were carried out by choosing 
specific initial bubble configurations. In order to study this mechanism 
in a realistic situation of a phase transition, we carried out a full 
study of the effectiveness of this mechanism, for the case of zero 
explicit symmetry breaking, by simulating a first order transition via 
random nucleation of bubbles \cite{dgl3}. We showed there that for very 
low bubble nucleation rates (leading to large inter-bubble separation 
and hence very energetic collisions), the new mechanism becomes
the dominant mode of production of well separated vortices. However, for 
large nucleation rates the vortex-antivortex pairs produced via this 
mechanism consist of strongly overlapping field configurations which 
decay rapidly. Thus for large nucleation rates, number of surviving
vortices will be roughly the same as expected from the 
Kibble mechanism. These results are
important as they show that defect production in certain situations (i.e. 
very low nucleation rate) may be drastically altered due to contributions 
of this mechanism. For example, this mechanism may completely
dominate the production of cosmic strings, monopoles etc. in models of 
extended inflation in the early Universe, and may play an important role
in the production of vortices in superfluid $^3$He A - B transition
\cite{volov}.  

  From the results in ref. \onlinecite{dgl1} we expect that there is a 
special class of systems, those with small explicit symmetry breaking, 
where this mechanism may be remarkably effective in defect production 
during a phase transition. There are many examples of systems with 
spontaneously broken symmetries where the symmetry is also broken 
explicitly. In the context of particle physics, the Skyrmion picture 
of baryons in the context of chiral models is an example where explicit 
symmetry breaking terms are needed to incorporate a non-zero pion mass.
In fact it was this system where the role of explicit symmetry breaking
in enhancing defect production was first discussed by Kapusta and one 
of the author \cite{kpst}, though the underlying mechanism was different
there. There is another example, in Particle physics, that of axionic
strings arising from the breaking of the Pecci-Quinn symmetry, where
explicit symmetry breaking is present \cite{axn}. It was suggested in 
\cite{dgl1,dgl2,dgl3} that this new mechanism may be especially
effective for the production of axionic strings due to the presence of
explicit symmetry breaking. However, that argument does not seem to be
correct since the scale of Pecci-Quinn symmetry
breaking is very large (of order $10^{10}$ GeV or so) compared to the 
scale of explicit symmetry breaking, which arises at the QCD scale. Thus,
at the Pecci-Quinn scale, explicit symmetry breaking can not play any role 
in the production of axionic strings, though this new mechanism may 
certainly contribute to their production.

 In condensed matter, nematic liquid crystals provide a simple example of 
such systems where the presence of external electric or magnetic fields 
induces explicit symmetry breaking terms \cite{lc}. Hence this new 
mechanism may play a dominant role there. Though there are subtleties in 
considering this system due to the fact that opposite orientations of the
order parameter are identified. We will discuss this point later in 
Section V.

  From the point of view of these systems, and in general for systems
with explicit symmetry breaking, it is important to know the actual
contribution of this mechanism in a realistic phase transition. The study 
in ref. \onlinecite{dgl1} considered some specific bubble configurations
and showed how a very large number of defects may be produced in a single 
two bubble collision. What one needs to know is the average number of 
defects produced via this mechanism when bubbles are randomly nucleated. 
We have already carried out such a study for the case of zero explicit 
symmetry breaking, which is the case of widest possible applications. 
[For example, the production of cosmic strings, magnetic monopoles etc., 
in the early Universe,  and the production of vortices in superfluid 
$^3$He and $^4$He in the laboratory.] As the detailed dynamics, and hence
the defect production via this mechanism, is very different for the 
systems with explicit symmetry breaking, we need to carry out detailed 
simulations of the phase transition for this case as well. Certainly, for 
baryon (Skyrmion) production in the heavy-ion collisions, as well as for 
the production of liquid crystal strings in the presence of external 
fields, the results of ref. \onlinecite{dgl3} (for the zero symmetry 
breaking case) are not much useful. We should expect that a much more 
dramatic role will be played by the new mechanism for these systems, 
even in the generic situation of a phase transition. 

 There is one more reason why this mechanism may be most dominant in 
the situation with explicit symmetry breaking. If bubble nucleation is 
not happening at very large temperatures (compared to the explicit 
symmetry breaking scale) then the distribution of phases inside 
different bubbles will get biased.  This bias will suppress vortex 
production due to Kibble mechanism. For example, in our simulation 
with 63 bubbles (discussed in Section IV), essentially all the
bubbles had $\theta$ which was either less than $\pi/2$ or greater
than $3\pi/2$ (for the case when $\theta = 0$ corresponds to the true 
vacuum).  Due to this, not a single vortex 
was produced via the Kibble mechanism.
Bias in $\theta$ towards value zero does have adverse effect even on 
the new mechanism due to reduced potential energy which affects the
flipping of the field. However, wall oscillations will be expected
to be much larger now due to smaller $\theta$ gradients between 
colliding bubbles. This should help vortex production via the new
mechanism. What we find is that all the vortices and antivortices
(a total of 104 in this case) were produced via the new mechanism.
This is about the same average vortex production per bubble via the
new mechanism as we obtain when $\theta$ distribution inside
bubbles is chosen to be uniform (though in that case there are 
some vortices produced via the Kibble mechanism as well). 

 We also would like to know how this average defect production is affected 
by dynamical details, such as the magnitude of explicit symmetry breaking, 
nucleation rate, presence of damping etc. We carry out a complete study of
all these issues in this paper. We find that defect production increases 
with increasing magnitude of explicit symmetry breaking. Defect production 
also increases for lower nucleation rates due to bubble collisions being 
more energetic. This is the same behavior which was observed in ref. 
\onlinecite{dgl3} for the case of zero explicit symmetry breaking.  
Presence of damping suppresses field oscillations and hence results in 
lower number of defects. However, compared to the case of zero explicit 
symmetry breaking discussed in ref. \onlinecite{dgl3} where strong 
damping makes this mechanism completely ineffective, for the present case 
we find that some pairs are always produced via this mechanism (as long 
as explicit symmetry breaking
is not too small).  This results is very significant for the cases of 
experimental interest.  For example, in the case of Skyrmion production in 
heavy-ion collisions, dissipative effects will always be present in the 
expanding plasma. This is especially true for the case of liquid crystal 
strings where the dynamics is completely dominated by dissipation.

 The paper is organized in the following manner. The second section 
reviews the essential physical picture of the new mechanism \cite{dgl2} 
and  discusses the numerical technique. Section III discusses the results 
of the simulation of the phase transition by random nucleation of 
bubbles with unbiased $\theta$ distribution inside bubbles. In Section
IV we present results for simulation with $\theta $ distribution inside
bubbles being biased according to the explicit symmetry breaking term.
The dependence of vortex production on nucleation rate and the magnitude 
of explicit symmetry breaking is discussed in Section V. The effect 
of damping on this mechanism is discussed in Section VI and conclusions
are presented in Section VII.  

\vskip .3in
\centerline{\bf II. PHYSICAL PICTURE OF THE MECHANISM}
\centerline{\bf AND NUMERICAL TECHNIQUE}
\vskip .1in
         
 We first briefly recall the essential features of this mechanism 
\cite{dgl2}. Consider the case of spontaneous breaking of a global U(1) 
symmetry, with the order parameter being the vacuum expectation value 
of a complex scalar field $\Phi$. The vacuum manifold is a circle $S^1$. 
After the phase transition, $\Phi$ will have a non-zero magnitude but 
its  phase $\theta$ may vary spatially. Consider a region of space 
where $\theta$ has some small variation between two points A and B. Now 
suppose that the magnitude of the field undergoes strong oscillations in 
a small region between points A and B. If $\Phi$ passes through zero 
magnitude while oscillating in a region, then it is easy to see that 
$\theta$ in that region will discontinuously change to $\theta + \pi$. 
We call it the flipping of $\Phi$. By drawing the distribution of 
$\theta$, with flipped orientation in the middle portion between A and 
B, one can see that a vortex and antivortex pair has formed in the 
opposite sides of the line joining A and B, see ref. \onlinecite{dgl2} 
for details. These considerations can easily be generalized to other
defects \cite{dgl2}.

 Numerical techniques used in this paper are the same as used in the 
earlier papers \cite{dgl1,dgl2,dgl3}. We study $2+1$ dimensional case,
with the Lagrangian taken as the following.

\begin{equation}
{\it L} = {1 \over 2} \partial_{\mu} \Phi^{\dag} \partial^{\mu} \Phi
- {1 \over 4} \phi^2 (\phi - 1)^2 +
\epsilon \phi^3 + \kappa \phi^2 cos \theta
\end{equation}

 This Lagrangian is written in terms of appropriately scaled coordinates,
and a dimensionless complex scalar field $\Phi$, with $\phi$ and $\theta$ 
being the magnitude and the phase of $\Phi$. Value of $\epsilon$
is taken to be 0.05. This Lagrangian describes a 
theory where U(1) global symmetry is spontaneously broken, except for the 
presence of last term which breaks U(1) explicitly.

      For the case of $\kappa=0$, the process of vortex creation via 
bubble nucleation has been studied in detail in ref. \onlinecite{ajit} 
for the Kibble mechanism, and in ref. \onlinecite{dgl2,dgl3} for the new 
mechanism. At zero temperature, the phase transition takes place via 
nucleation of bubbles of true vacuum in the background of false vacuum 
via quantum tunneling. The bubble profile $\phi$ is obtained by solving 
the Euclidean field equation \cite{bbl}  

\begin{equation}
{d^2 \phi \over dr^2} + {2 \over r} {d \phi \over dr} - 
V^\prime(\phi) = 0
\end{equation}

\noindent where $V(\phi)$ is the effective potential in Eq.(1) (with 
$\kappa = 0$) and $r$ is the radial coordinate in the Euclidean space. In 
the Minkowski space initial profile for the bubble is obtained by putting 
$t=0$ in the solution of the above equation. $\theta$ takes a constant 
value inside a given bubble.

 The bubble profile found by solving Eucledian equation of motion Eq.(2) 
for $\kappa=0$ provides an adequate starting bubble configuration even 
for small non-zero values of $\kappa$ as we are only interested in 
vortex formation which happens after bubbles expand and coalesce. This 
was the approach used in our earlier work \cite{dgl1} for the case with 
$\kappa \ne 0$. However, there is a nontrivial issue here. [This was 
pointed out to us by Copeland and Saffin]. The standard method of 
finding the bounce solution \cite{bbl}, for the present case with 
$\kappa \ne 0$, works fine if the phase $\theta$ takes value 0 or $\pi$ 
inside the bubble. For any other choice of $\theta$, the method does not 
work with the conventional approach. It appears that if at all there is 
any bounce solution for arbitrary $\theta$, it must
require a non-uniform $\theta$ inside the bubble \cite{bbl2}. 

 For the purpose of the present paper, we will ignore this issue. One
approach can be that if one was considering thermal nucleation of 
bubbles, then if the explicit symmetry breaking term is very small 
compared to the scale set by the temperature, then one can ignore it 
while considering the nucleation process itself. Thus we will continue 
to use the bubble profile obtained for the case of $\kappa = 0$ and 
evolve the configuration with non-zero value of $\kappa$. The profile 
of the bubble calculated with $\kappa=0$ has the asymptotic value 
$\phi=0$ which is the local minimum of the effective potential. Hence 
it is suitable for nucleating the bubbles 
in the background of false vacuum, $\phi=0$. This suggests that
the explicit symmetry breaking term should be chosen such that the 
false vacuum remains the same $\phi=0$ so that there is no mismatch
between the false vacuum and the asymptotic value  of $\Phi$ of
different bubbles. The specific form of the explicit symmetry breaking 
term we use is motivated by this consideration as well as by simplicity. 
It is important to mention that our results are not sensitive to the form 
of the explicit symmetry breaking term, as long as the true vacuum (with 
non-zero value of $\kappa$) is non-degenerate.  We have verified that 
similar enhancement in vortex production results with other types of 
symmetry breaking terms as well, for example with $\kappa\phi cos\theta$.

  Even though we are neglecting the effects of explicit symmetry 
breaking in solving Eqn.(2) for the bubble profile, its effects should
be taken into account when $\theta$ inside bubbles is being determined
($\theta$ inside a given bubble taken to be uniform). For $\kappa = 0$ 
one takes all values of $\theta$ to be equally probable. However,
with $\kappa \ne 0$ one will expect a bias in $\theta$ distribution
inside bubbles as bubbles with smaller values of $\theta$ will have
smaller free energy and hence larger nucleation probability. We take
into account this bias in $\theta$ distribution inside bubbles later
in Section IV. We show there that, in a simulation involving nucleation 
of 63 bubbles, bias in $\theta$ distribution leads to all bubbles
having $\theta$ configurations such that not a single Kibble 
vortex/antivortex is produced. At the same time vortex production
due to the new mechanism is essentially unaffected by the biased
$\theta$ distribution. However, in order to have a conservative
estimate of the importance of the new mechanism compared to 
the Kibble mechanism, we first study situation where bubbles are
nucleated with uniform probability of $\theta$ for which Kibble 
mechanism contribution is well understood. This approximation will be 
consistent when the tilt is small compared to the temperature scale
at which bubble nucleation is supposed to be happening. [Though,
in that situation one should worry about validity of classical evolution
using field equations as for large temperatures vortex-antivortex
pairs can be thermally nucleated. We will consider this point later
in Section IV. For now we will continue using uniform probability
distribution for $\theta$ inside bubbles.]

 Rest of the numerical technique is the same as used in earlier works
\cite{dgl1,dgl2,dgl3}. The field evolution is carried out using field 
equations obtained from Eqn.(1). The simulation of the phase transition 
is carried out by nucleating bubbles on a square lattice with periodic 
boundary conditions, i.e on a torus. Lattice spacing in spatial 
directions was taken to be $\bigtriangleup x=0.16$, with time 
step $\bigtriangleup t$ equal to $\bigtriangleup x/\sqrt{2}$.
Simulations were carried out on an HP-735 workstation, a Silicon Graphics
Indigo 2 workstation, and an HPK-260 workstation at the Institute of 
Physics, Bhubaneswar.

When evolved with these equations, bubbles grow, coalesce, and vortices 
are formed in the intersection region of three or more bubbles by Kibble 
Mechanism, as well as  by the decay of bubble walls via the new mechanism. 
Note that, when $\kappa$ is non zero then $\theta=0$ is energetically 
preferred (for our choice of explicit symmetry breaking term).  So when a 
bubble with non-zero $\theta$ is nucleated, $\theta$ inside the bubble 
evolves towards $\theta=0$ and eventually starts oscillating about 
$\theta=0$ with decreasing amplitude. This non-trivial dynamics of 
$\theta$ is one feature which makes the present case qualitatively 
different from the case of $\kappa = 0$ considered in ref. 
\onlinecite{dgl3}. We explain this in some detail in the following
by considering a two bubble collision. 

  When two bubbles collide there is always $\phi$ 
oscillations in the coalesced portion of the two bubbles. Magnitude of 
this oscillation depends on the $\theta$ difference between the two 
bubbles as well as on the spatial  separation between the two bubbles. 
As described in ref. \onlinecite{ajit}, $\phi$ oscillations are more
prominent when $\theta$ difference between the bubbles is small.
Large $\theta$ difference leads to large gradient energy in the 
coalesced region which suppresses $\phi$ oscillations in that region.
When $\phi$ oscillations have sufficient amplitude then $\Phi$ can
overshoot the value $\Phi = 0$ by climbing over the potential hill.
Clearly, this leads to a change in $\theta$ in that region to $\theta
+ \pi$, i.e. to flipping of $\Phi$. If this oscillation, and the 
associated flip, happens inside a region where $\theta$ is spatially
varying, then a pair of vortex-antivortex forms in this single flip
(as explained at the beginning of this section). Note that so far in 
this picture we did not use anywhere the presence of explicit symmetry 
breaking (or, the rolling of $\theta$ to zero). One will expect that 
every time this type of situation arises (even in the course of a single 
two bubble collision), it will lead to a vortex-antivortex pair getting 
nucleated. In fact, this is where the explicit symmetry breaking term in 
the effective potential plays a crucial role, as we describe below. 
   
  Consider two bubbles with associated values of $\theta$ being 
$\pi + \alpha$ and $\pi - \alpha$ respectively, with $\alpha < \pi/2$. 
When these two bubbles collide, according to geodesic rule, $\theta$ in 
the coalesced portion takes the value $\theta = \pi$ (for this $\alpha$
should be small as $\theta$ rotates towards zero). Since $\theta$
is rolling towards zero in both the bubbles we always encounter
$\theta=\pi$ as we go from one bubble to the other and there
is a $\theta$ gradient which keeps  increasing due to rolling of 
$\theta$. It helps to think in terms of an arc spreading around
the value $\pi$ in the plot of the effective potential (see Fig.1)
which represents the variation of $\theta$ from the center of one 
bubble to the center of second bubble. Both ends of this arc (p and q) 
will roll down towards zero, and as the arc is constrained to go through
the value $\pi$ (due to continuity), it will lead to larger $\theta$ 
gradients. Now if $\Phi$ flips in this region due to $\phi$ oscillation 
then we will get a vortex-antivortex pair. The vortex and the antivortex 
in this pair move away from each other as $\theta$ is zero in between 
them while $\theta$ in the outer directions is $\pi$ having larger 
potential energy. This is how the first pair is created.

\begin{figure}[h]
\begin{center}
\vskip 0.2 in
\leavevmode
\epsfysize=5.0truecm \vbox{\epsfbox{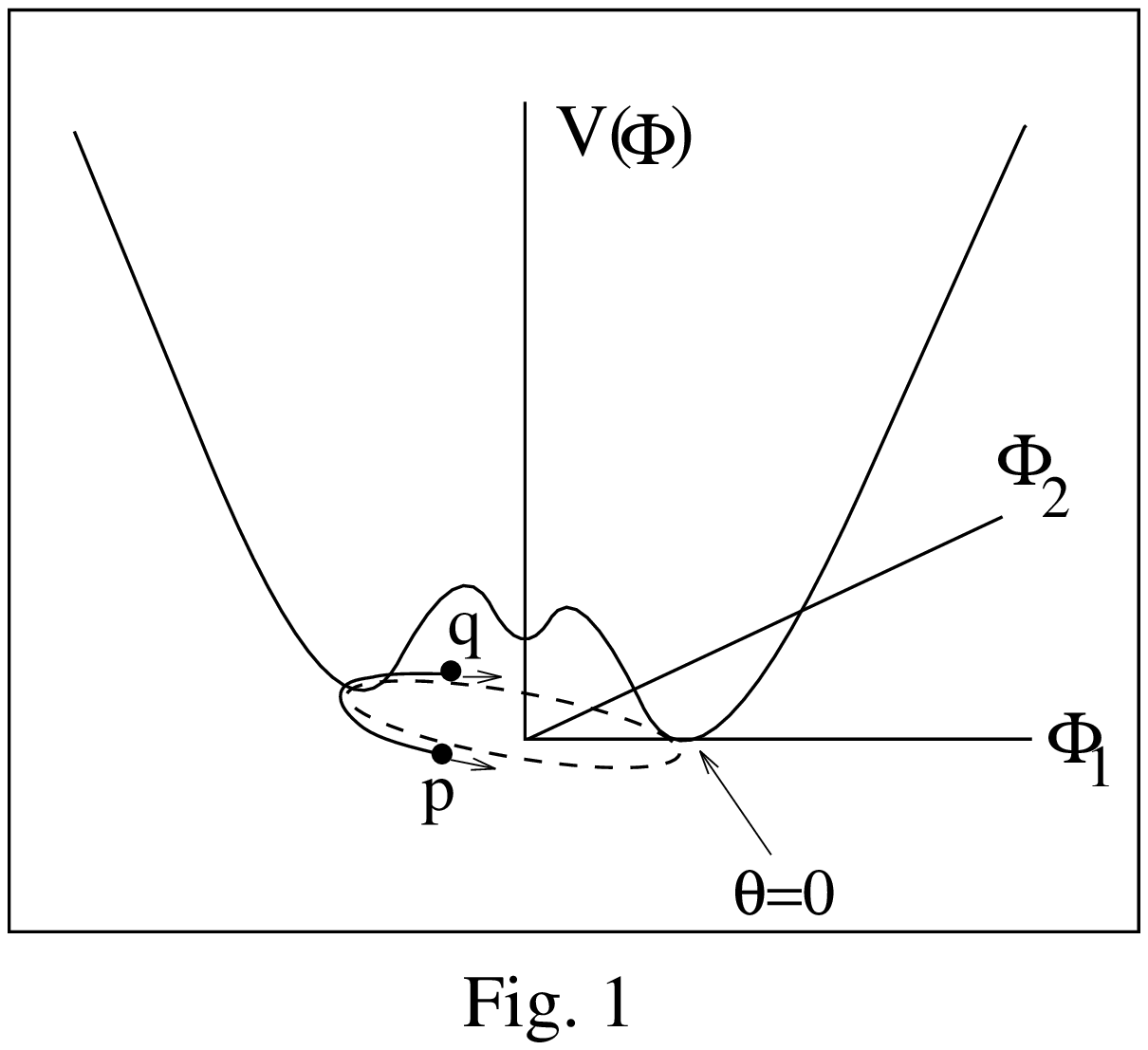}}
\vskip 0.2in
\end{center}
\caption{Plot of the effective potential. $\theta$ variation in
the physical space is taken to correspond to the solid arc. The
end points, p and q, of the arc roll towards $\theta = 0$, but the 
arc is constrained to go through $\theta = \pi$ due to continuity.}
\label{Fig.1} 
\end{figure}

  We note now that even though $\theta$ is $0$ in the coalesced 
portion, there is still a $\theta$ gradient which arises due to
$\theta$ oscillation around zero in the bubbles. [$\theta$ in
each bubble, while rolling down towards $0$, overshoots leading
to non zero $\theta$ gradient in that region.] If $\phi$
oscillations are still strong enough so that they lead to yet
another flip of $\Phi$ then another vortex-antivortex pair
will be nucleated.  This process of successive pair creations
continues till we have $\theta$ oscillations maintaining $\theta$ 
gradient in the coalesced portion and large $\phi$ oscillations 
flipping $\Phi$. Creation of multiple pairs in this manner was
demonstrated in ref. \onlinecite{dgl1}, where it was shown that
as many as ten vortices and antivortices were  created in a single,
two bubble collision. 

We have emphasized that if there is no flipping of $\Phi$
then there will be no pair creation of vortices.  Also, the
amplitude of  $\phi$ oscillation in the coalesced region
depends on the separation of bubbles, with larger separation leading 
to larger amplitude.  This may lead one to think that if the bubbles 
are at the closest separation then there will be no large oscillation 
of $\phi$ so there may not be any pair creation at all. This was indeed
true for the case of $\kappa = 0$ discussed in ref. \onlinecite{dgl3}.
However, this is not true for the present case, with $\kappa \ne 0$.
Consider for example a configuration which leads to $\theta=\pi$ in the 
coalesced region. The gradient energy density in $\theta$ variation 
in the coalesced region will keep increasing as $\theta$ in both 
the bubbles rolls towards $0$.  This, implies large 
energy in the coalesced portion due to explicit symmetry breaking
term which can drive $\phi$ to climb the potential hill and lead
to the flipping of $\Phi$. [At least when $\kappa$ is not too small,
see below.] Therefore, there is at least one
pair of vortex-antivortex pair nucleated for appropriate initial 
$\theta$ even for smallest separation of the bubbles.  As we will see 
later, even in the presence of strong damping one pair is still nucleated 
(in contrast to the case of ref. \onlinecite{dgl3}), though successive 
nucleation may be suppressed depending on the magnitude of damping.
 
\vskip .3in
\centerline{\bf III. SIMULATION OF THE PHASE TRANSITION}
\vskip .1in

    The preceding section explains how vortex-antivortex pairs
are created in a two bubble collision in the presence of explicit
symmetry breaking. It was shown in ref. \onlinecite{dgl1} that a large
number of vortex-antivortex pairs can form in a two bubble collision
for appropriately chosen initial field configuration of  bubbles.
Clearly, in an actual phase transition, such large enhancement
may not be expected due to randomness in values of $\theta$ inside
the bubbles as well as in their separations. This is especially
so because the vortex production via this mechanism depends on the
dynamical details of the bubble collisions and depends on parameters
such as the magnitude of symmetry breaking term in the effective 
potential. Therefore, to find actual enhancement in the vortex 
production in this case one has to carry out the full simulation of the 
phase transition. We describe the results of such simulations in
this section.

  In this section we will present results with the value of $\kappa =
0.015$. Here, we will neglect the explicit symmetry breaking term in 
determining $\theta$ inside nucleated bubbles. In the next section we 
will take into account the effect of explicit symmetry breaking in
biasing the $\theta$ distribution. We will see that the average defect
production per bubble, via the new mechanism, remains almost unaffected, 
while defect production via the Kibble mechanism becomes strongly 
suppressed for biased $\theta$ distribution case.
 
 Nucleation rate was suitably chosen so that there were all together 
seven bubbles nucleated in the whole lattice. Physical size of the
lattice was taken to be 192 $\times$ 192.  As we mentioned 
earlier, the location of bubbles in the lattice 
as well as the values of $\theta$ inside them were chosen randomly.  
Number of bubbles was chosen to be small (by taking small nucleation 
rate) so that bubble collisions are energetic which helps in 
vortex-antivortex pair creation.  Also the bubbles are nucleated only 
in a short time duration so they are roughly of same size when they 
collide.  These bubbles then expand and coalesce. 

Fig.2a shows the plot of $\Phi$ at an early stage when bubbles have just 
started coalescing.  [In all the plots of $\Phi$, the orientation of an 
arrow from the positive x axis corresponds to $\theta$ and the length of 
the arrow is proportional to $\phi$.] Rotation of $\theta$ can be clearly 
seen inside the bubbles. Due to $\phi^2$ dependence in the $\kappa$ term, 
$\theta$ rotation is slowest near the bubble walls. Note that $\theta$ in 
the central region of many bubbles has almost rotated to zero, the 
only memory of the initial values of $\theta$ remaining near the walls. 
Interestingly, distribution of $\theta$ at walls of different bubbles 
plays the most dominant role in the production of all the vortices 
observed in the simulations.

 The location of the vortices was determined by using an algorithm to
locate the winding number. As the phase transition nears completion
via the coalescence of bubbles, magnitude of $\Phi$ becomes non-zero
in most of the region with well defined phase $\theta$. We divide each 
plaquette in terms of two (right angle) triangles and check, for each 
such triangle, whether a non-zero winding is enclosed. For this purpose
we use the geodesic rule to determine $\theta$ configuration in between
two adjacent lattice points, see ref.\onlinecite{dgl3} for details. 
Windings are detected only in regions where the magnitude of $\Phi$
is not too small in a small neighborhood of the triangle under
consideration. If $\Phi$ is too close to zero in a region then
that region is still mostly in the false vacuum and there is no
stability for any windings present there. After getting {\it probable} 
locations of vortices using the above algorithm, we check the region
containing each vortex/antivortex, using detailed phase plots and 
surface plots of $\phi$ to check the winding of the vortex, and
select only those vortices which have well defined structure. 
By checking similar plots at earlier as well 
as later time steps we determine whether the vortex was produced
due to oscillation, and subsequent flipping,  of $\Phi$, or via
the Kibble mechanism.

\begin{figure}[h]
\begin{center}
\vskip -0.5 in
\leavevmode
\epsfysize=10truecm \vbox{\epsfbox{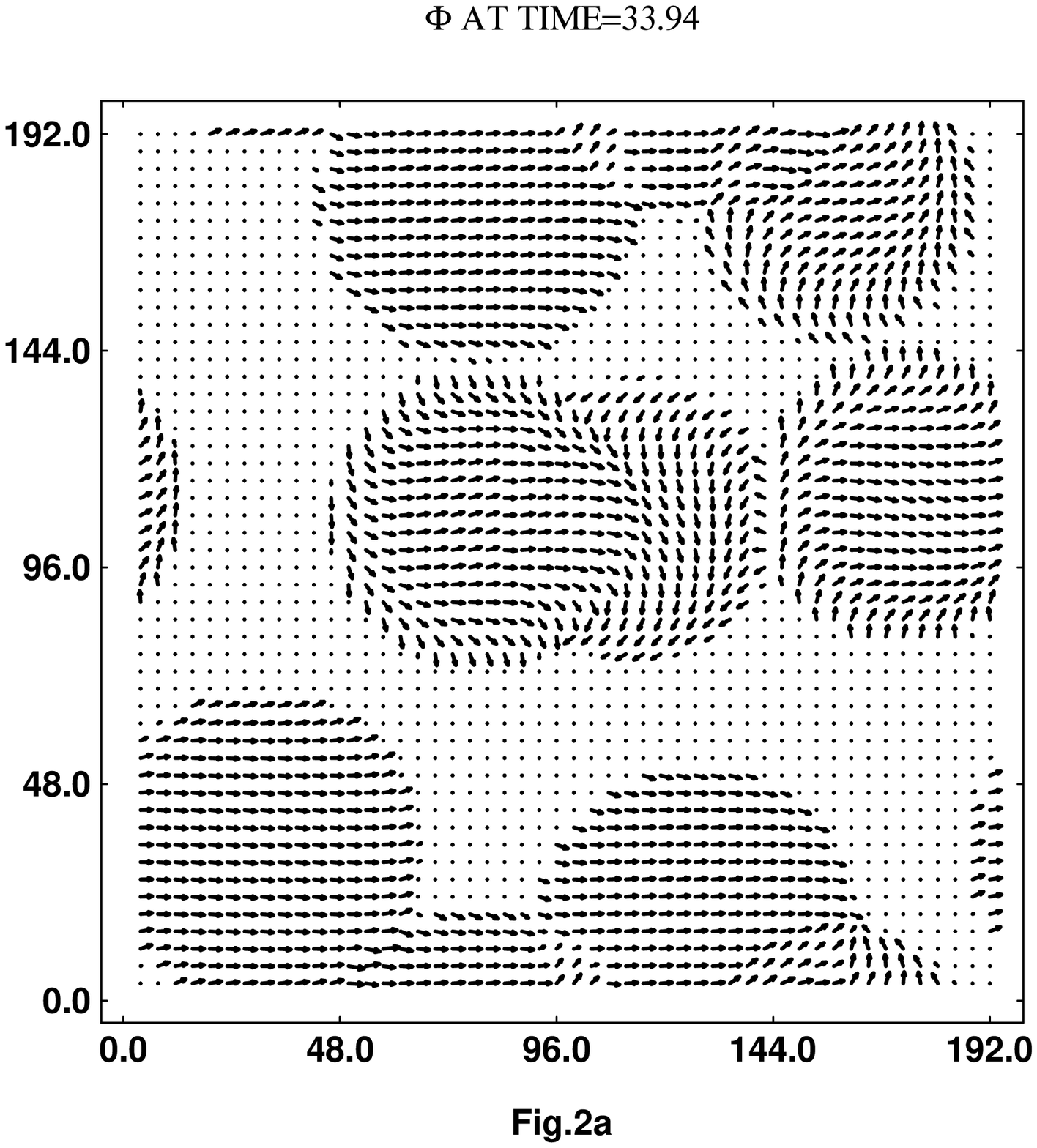}}
\vskip -0.5in
\end{center}
\end{figure}

\begin{figure}[h]
\begin{center}
\vskip -0.5 in
\leavevmode
\epsfysize=10truecm \vbox{\epsfbox{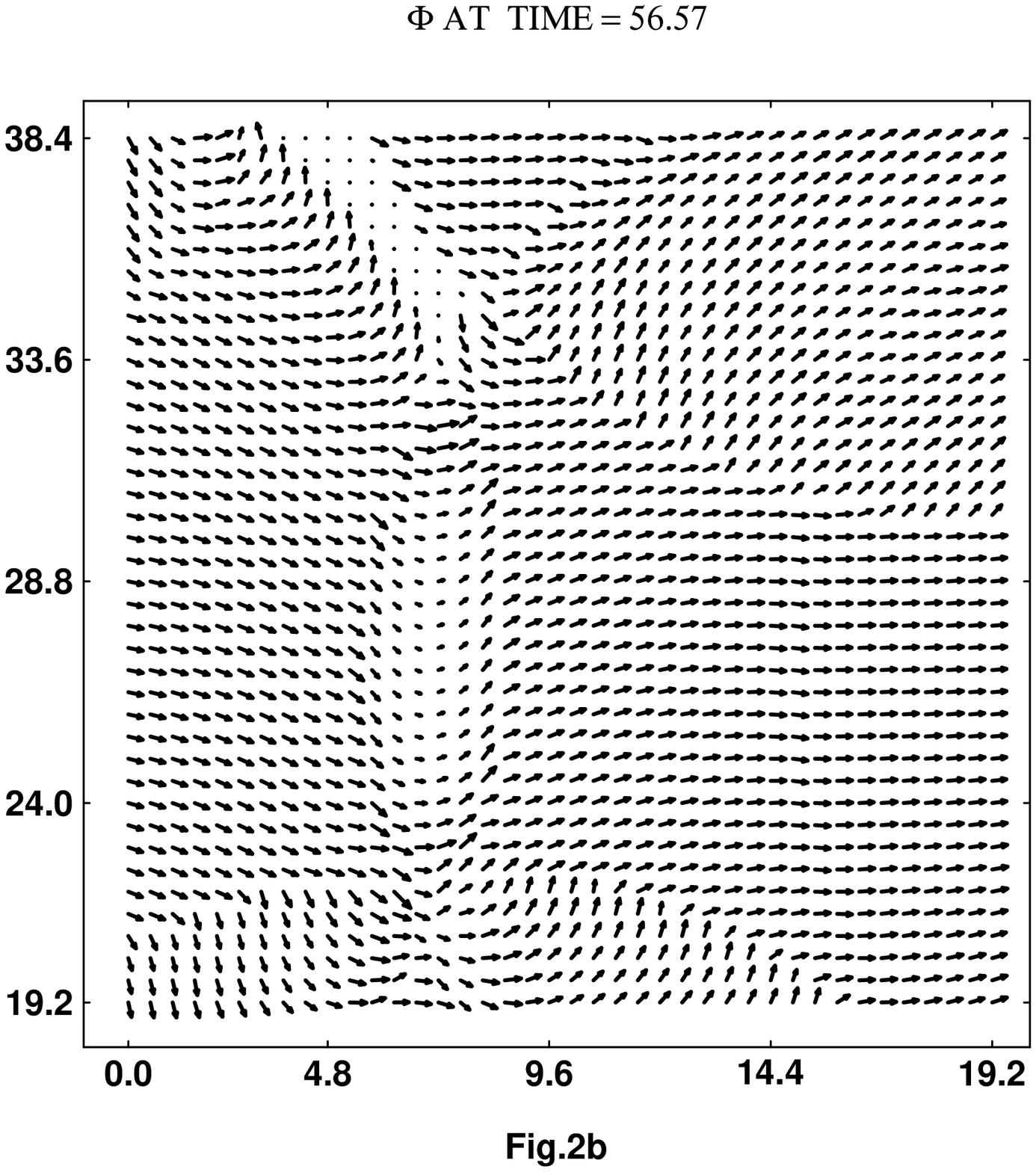}}
\vskip -0.5in
\end{center}
\end{figure}

\begin{figure}[h]
\begin{center}
\vskip -1.0 in
\leavevmode
\epsfysize=10truecm \vbox{\epsfbox{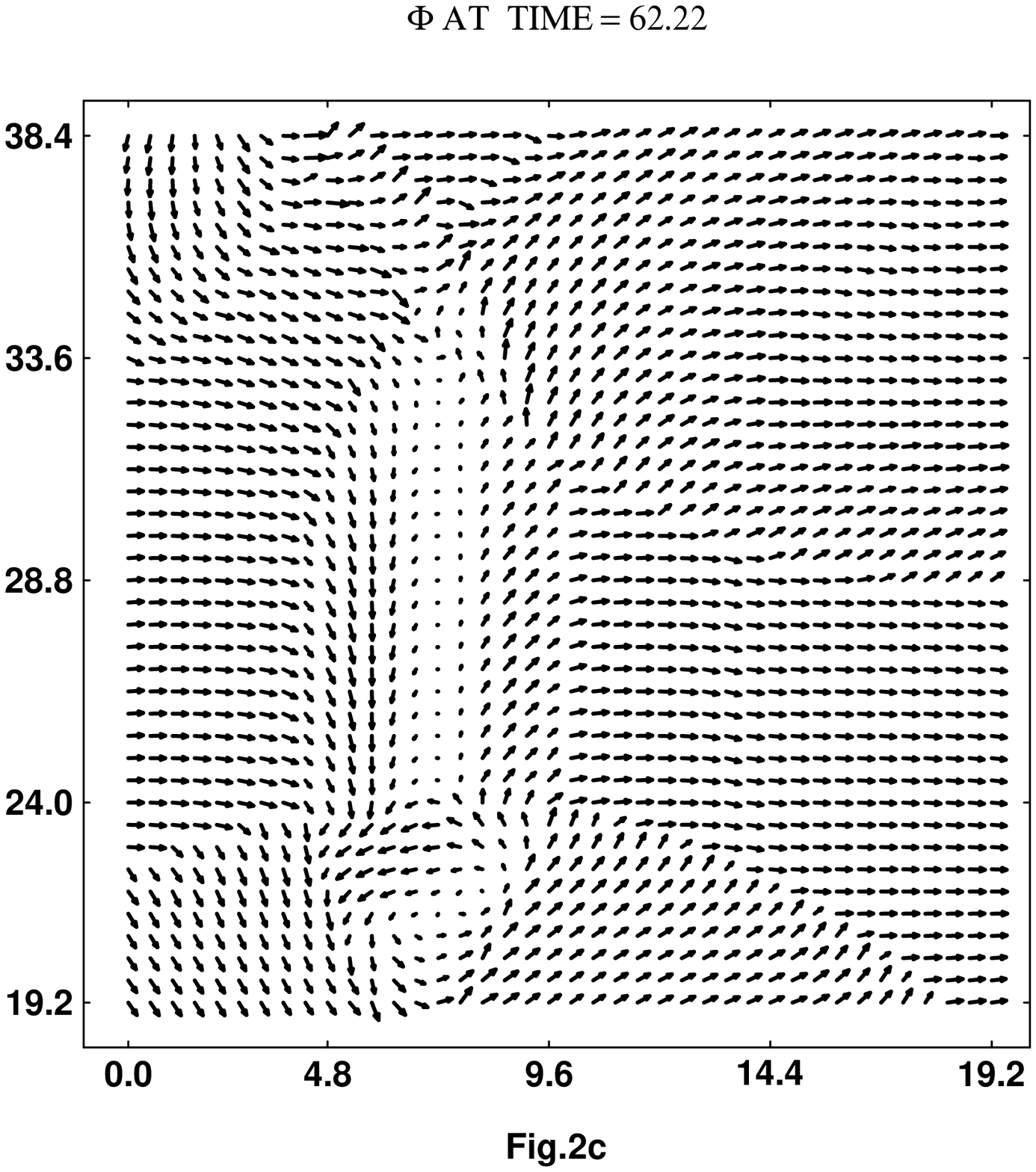}}
\vskip -0.5in
\end{center}
\end{figure}

\begin{figure}[h]
\begin{center}
\vskip -1.0 in
\leavevmode
\epsfysize=10truecm \vbox{\epsfbox{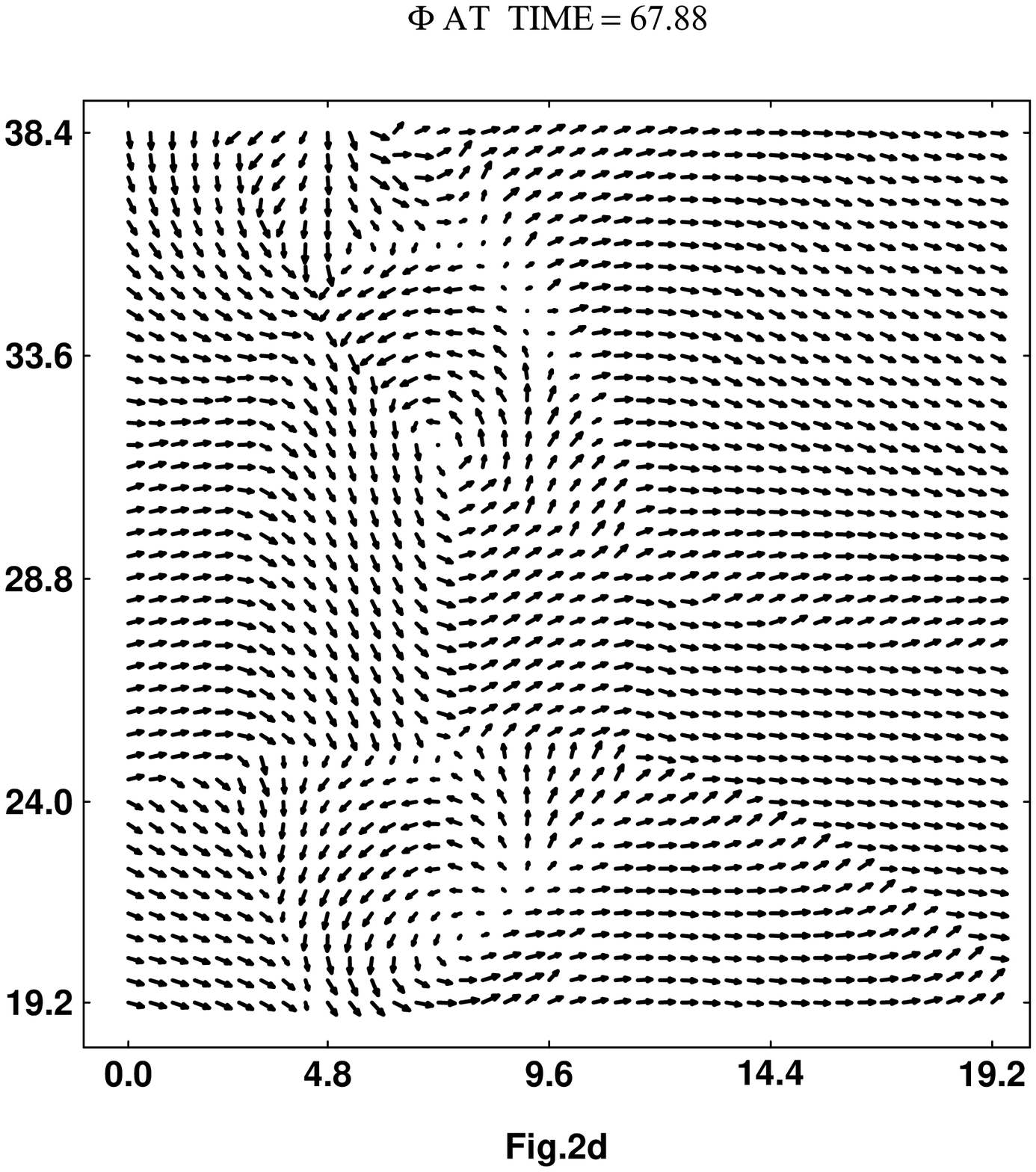}}
\vskip -0.5in
\end{center}
\end{figure}

\begin{figure}[h]
\begin{center}
\vskip -1.0 in
\leavevmode
\epsfysize=10truecm \vbox{\epsfbox{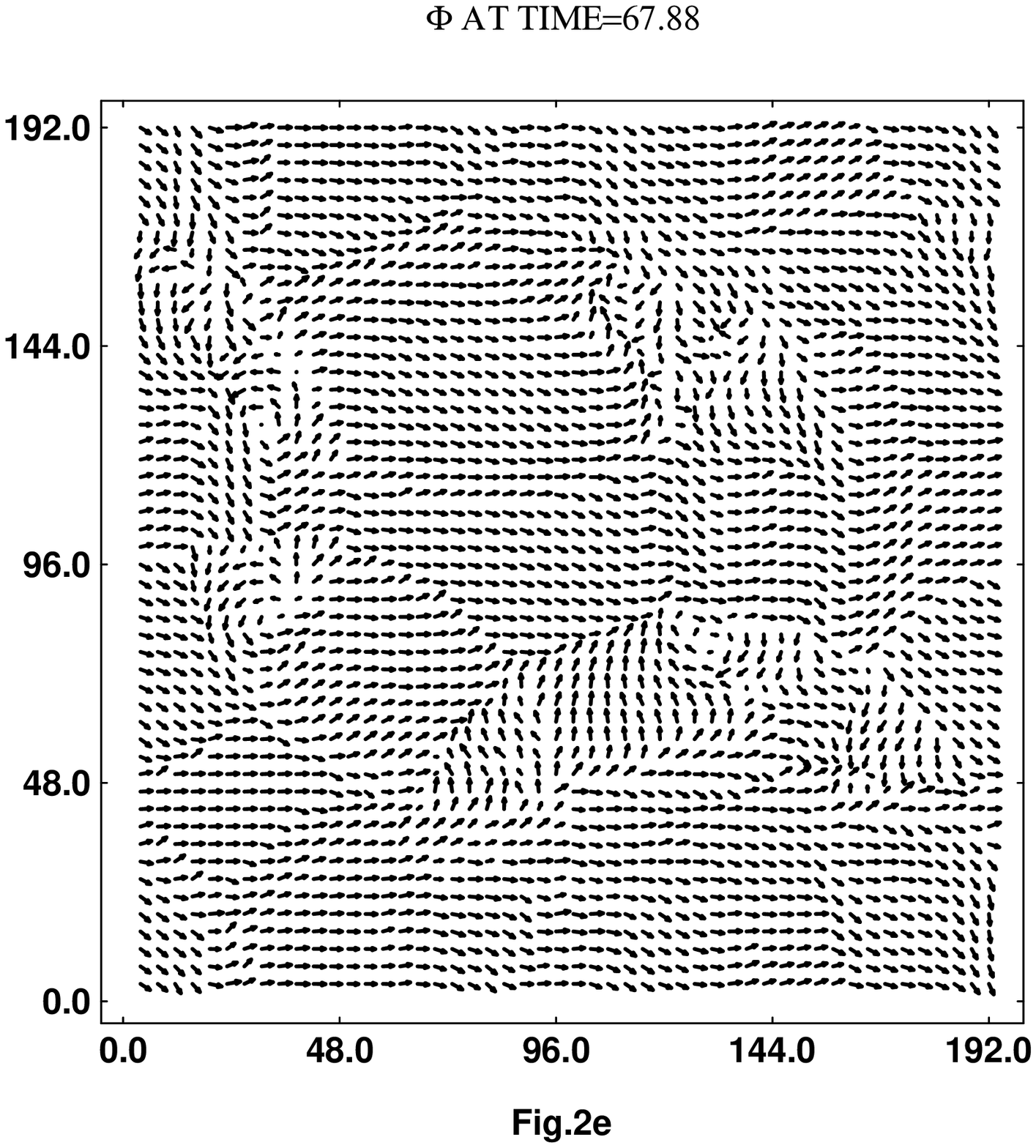}}
\end{center}
\caption{(a) Plot of $\Phi$ at an early stage when bubbles have just 
started coalescing. In all the plots of $\Phi$, the orientation of an 
arrow from the positive x axis corresponds to $\theta$ and the length 
of the arrow is proportional to $\phi$. (b) $\Phi$ plot in a region 
where there is no winding present at this stage. (c) An elongated region 
towards left of the center is oscillating, and a pair has been created. 
(d) Another pair has been created in this region due to continued 
oscillations.  (e) $\Phi$ plot of the entire lattice at an stage when 
maximum number of vortices and antivortices are present. Altogether, 
there are four vortex-antivortex pairs present at this stage.}
\label{Fig.2} 
\end{figure}
 
Figs.2b, 2c and 2d show a region where two vortex-antivortex pairs form 
due to field oscillations. Fig.2b shows the plot of $\Phi$ at $t = 
56.57$. There is no winding present in the region at this stage. Fig.2c 
shows the plot at $t = 62.22$. We see that an elongated region towards 
left of the center is oscillating, and a pair has been created. These 
oscillations continue, leading to creation of another pair, as shown in 
Fig.2d. Fig.2e shows the $\Phi$ plot of the entire lattice at this stage 
which corresponds to the situation when maximum number of vortices and 
antivortices are present. There are four vortex-antivortex pairs present 
at this stage. One of the pairs is near x = 36, y = 96, second pair is 
near x = 36, y = 132, third pair is near 
x = 136, y = 72, and fourth pair is near x = 116, y = 144.   
         
  In this particular simulation a total of (time integrated) 14 vortices 
and antivortices were formed. Out of these,  only two are formed due
to Kibble mechanism. This is consistent with the expected number density 
of 1/4 per bubble (for U(1) vortices in 2 spatial dimensions).  Remaining 
twelve vortices and antivortices are formed due to this new mechanism of 
coalesced bubble wall decaying by pair creation. This gives the average 
number of defects per bubble via the new mechanism to be about 1.7. 
Thus in this case this mechanism is completely dominant over 
the Kibble mechanism. Average number of all 
defects per bubble in this case is then 2.0 which is eight times larger 
than the number expected if vortices were only produced due to the Kibble 
mechanism.  In all our simulations we have found strong enhancement of 
the number density of defects per bubble in the presence of explicit 
symmetry breaking.   We should mention that apart 
from these 14 vortices and antivortices there 
were four additional pairs (i.e. eight vortices and antivortices) formed
where vortices and antivortices were not well separated. Still windings 
of $\theta$ corresponding to these pairs could be clearly seen 
in $\theta$ plots. We do not count such pairs as these defects annihilate
quickly. Such strongly overlapping pairs were found in all simulations.

 There were some intriguing features observed in these simulations
which we describe in the following. Fig.3a shows the field configuration 
at $t = 42.42$, where we see a well separated vortex-antivortex pair.
The vortex and the antivortex move towards each other, as shown in Fig.3b.
Due to asymmetric field configuration, the motion of the vortex-antivortex
pair is not along the direction of their separation, rather the pair moves
upward as a whole. The pair has annihilated by $t = 50.91$, as shown by
the absence of any windings in Fig.3c. Subsequent plot at $t = 53.74$ in
Fig.3d shows that the vortex-antivortex pair has been re-created. 
Interestingly, however, this time the location of the vortex and the 
antivortex are reversed. It is as if the vortex and the antivortex have 
passed through each other. Also, by checking subsequent plots, we find
that the motion of the pair as a whole is now downward. This new pair  
annihilates later. This pair, thus, represents a bound state of a 
vortex-antivortex system. There were several vortex-antivortex pairs
for which this cycle of annihilation and re-creation was seen. In all
the cases,  the vortex and the antivortex were exchanged after first 
annihilation and re-creation. For some cases, the cycle was repeated,
but when the pair was re-created second time, the vortex and the 
antivortex scattered back, instead of going through each other.

 This process of the vortex and the antivortex passing through each other
in the first cycle of annihilation and re-creation should be contrasted
with the results in ref. \onlinecite{carl,mmr} for vortex-antivortex
scattering. For global defects it was argued in \cite{carl} 
that for vortex-antivortex collisions which are dominated by gradient 
energy considerations, the vortex and the antivortex should retrace 
their original paths (i.e. they should bounce back) in a process of 
annihilation in a head-on collision and subsequent re-creation. Numerical 
simulation of gauged vortices in \cite{mmr} showed that in such 
collisions the vortex and the antivortex bounce back 
as long as their kinetic energies are not very large, but they
go through each other for extremely energetic collisions. In 
light of these studies, our numerical results of passing through of 
vortex and antivortex suggests that the collision is sufficiently 
energetic that gradient energy considerations are not relevant any more, 
like the results in ref. \onlinecite{mmr} for large kinetic energies. 
An estimate of the velocities from Figs.3a and 3b shows the velocity of 
the vortex to be between 0.9 and 1 and that of the antivortex, 
between 0.8 and 1. The uncertainty in the velocity arises from the 
uncertainty in the locations of the centers of the vortices. These are 
indeed very large velocities. One would expect such large velocities 
to arise here due to the domain wall (arising due to the concentration of 
$\theta$ gradient) stretching between the vortex and the antivortex. 
Scattering back of vortices in subsequent annihilations and re-creations
is then understood as due to velocities of vortices being not too large,
so the arguments of ref. \onlinecite{carl} should be applicable. This 
scattering back of vortex-antivortex thus provides an illustration of
the physical arguments given in ref. \onlinecite{carl}. We mention here 
that vortex-antivortex annihilations were seen in $\kappa = 0 $ case 
also \cite{dgl3}, but the vortex-antivortex pair was never re-created 
there. Presumably, the velocities at annihilation
are much larger for the present case of $\kappa \ne 0$.

\begin{figure}[h]
\begin{center}
\vskip -0.5 in
\leavevmode
\epsfysize=10truecm \vbox{\epsfbox{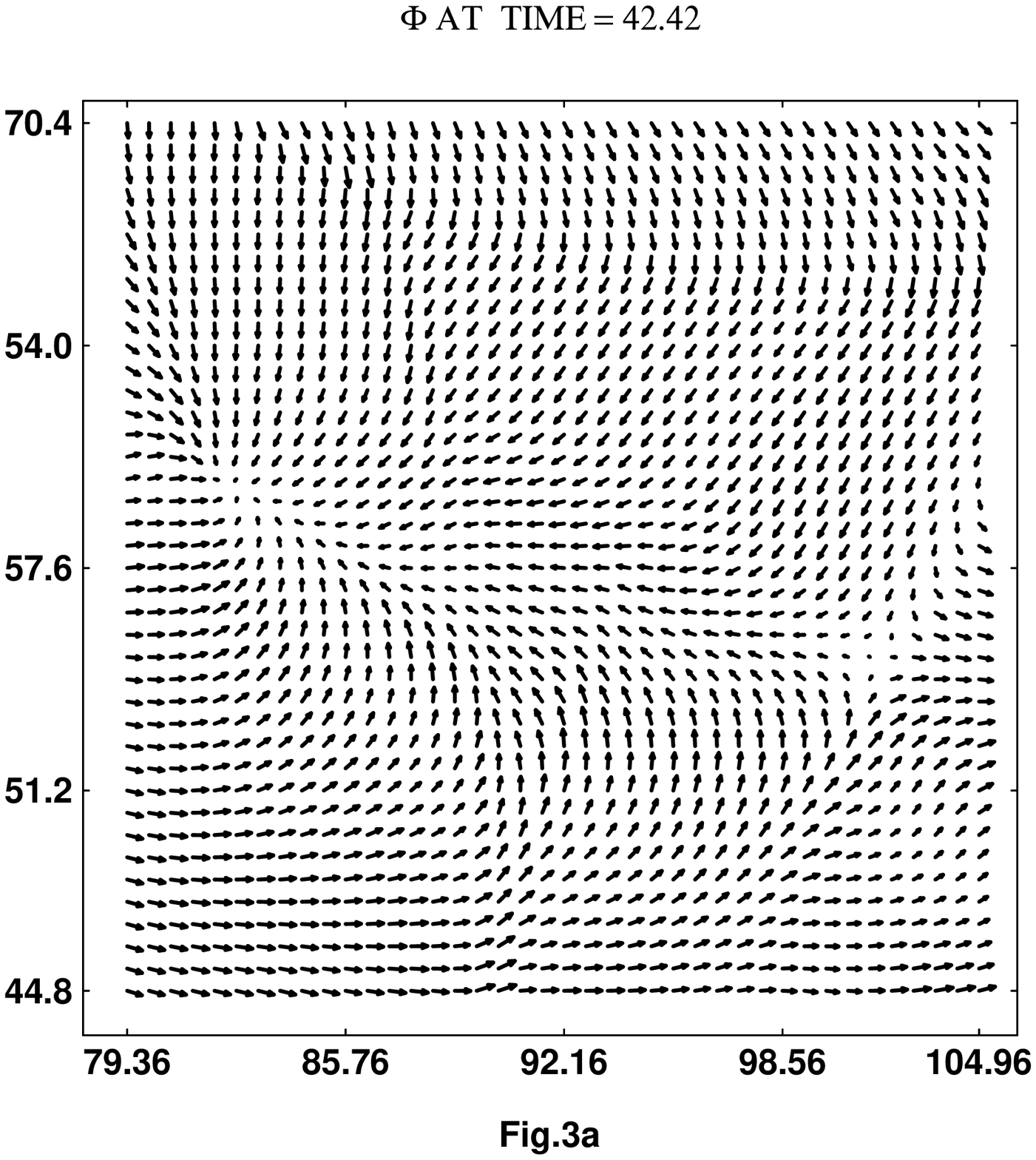}}
\vskip -0.5in
\end{center}
\end{figure}

\begin{figure}[h]
\begin{center}
\vskip -1.0 in
\leavevmode
\epsfysize=10truecm \vbox{\epsfbox{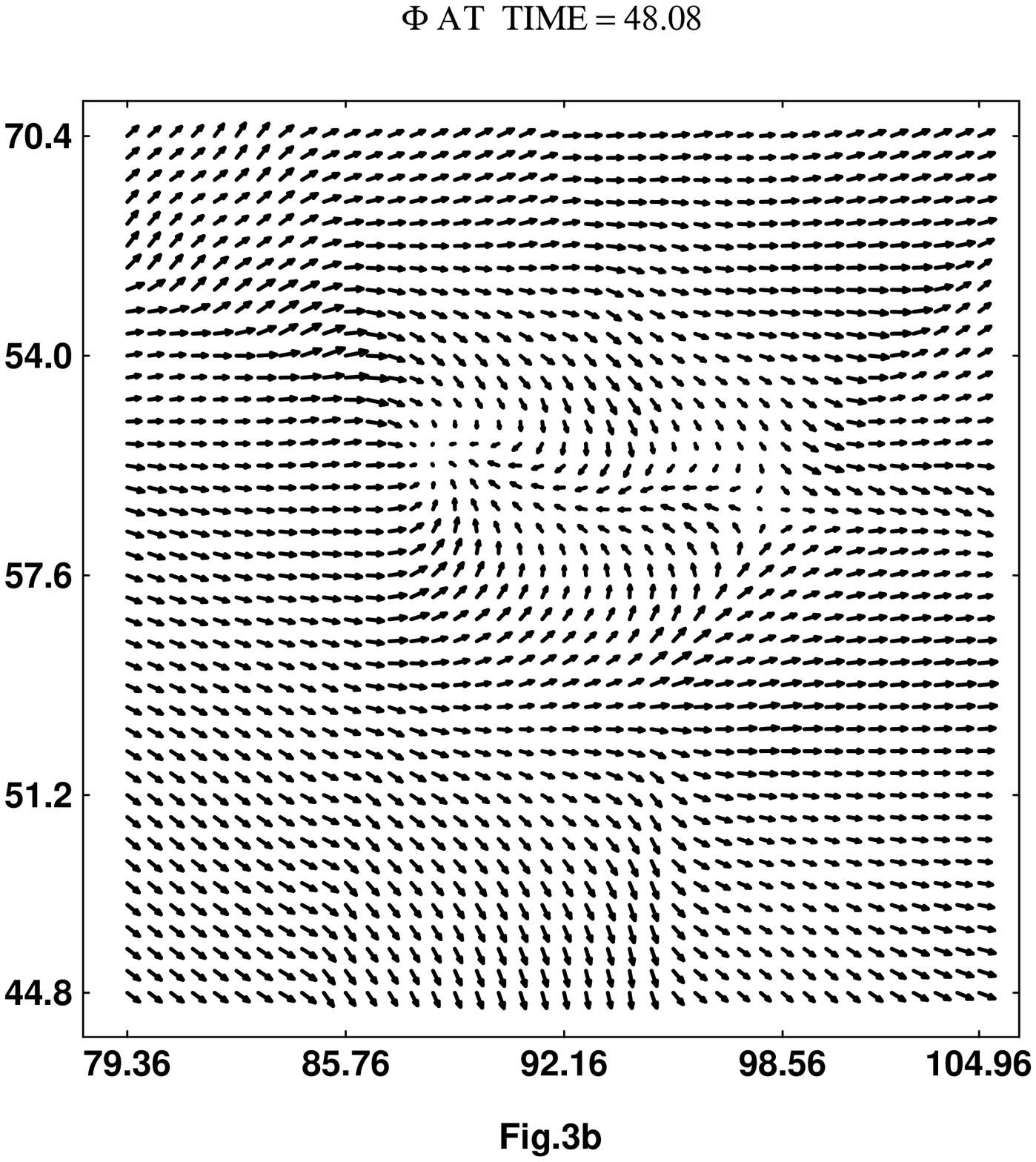}}
\vskip -0.5in
\end{center}
\end{figure}

\begin{figure}[h]
\begin{center}
\vskip -1.0 in
\leavevmode
\epsfysize=10truecm \vbox{\epsfbox{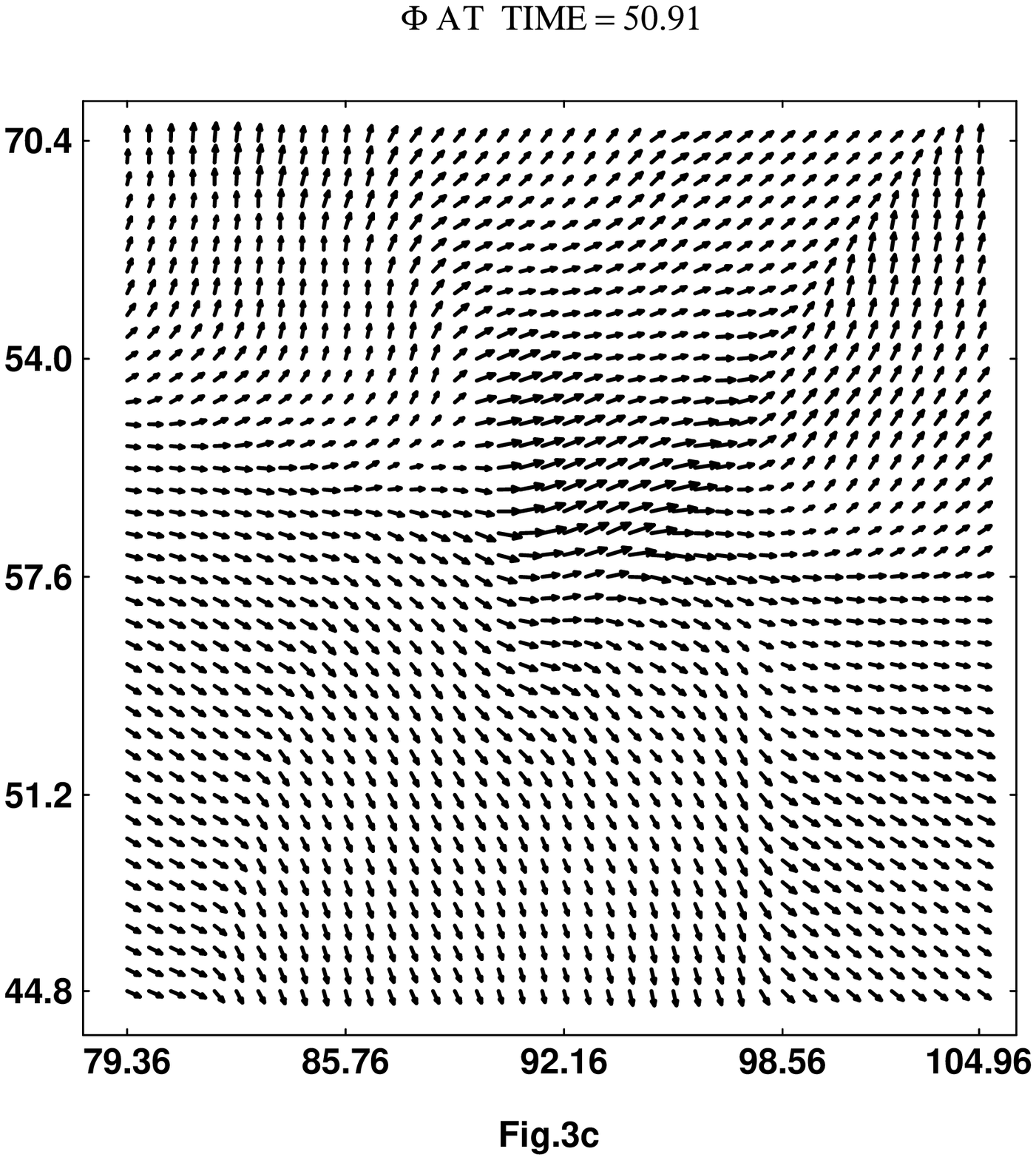}}
\vskip -0.5in
\end{center}
\end{figure}

\begin{figure}[h]
\begin{center}
\vskip -1.0 in
\leavevmode
\epsfysize=10truecm \vbox{\epsfbox{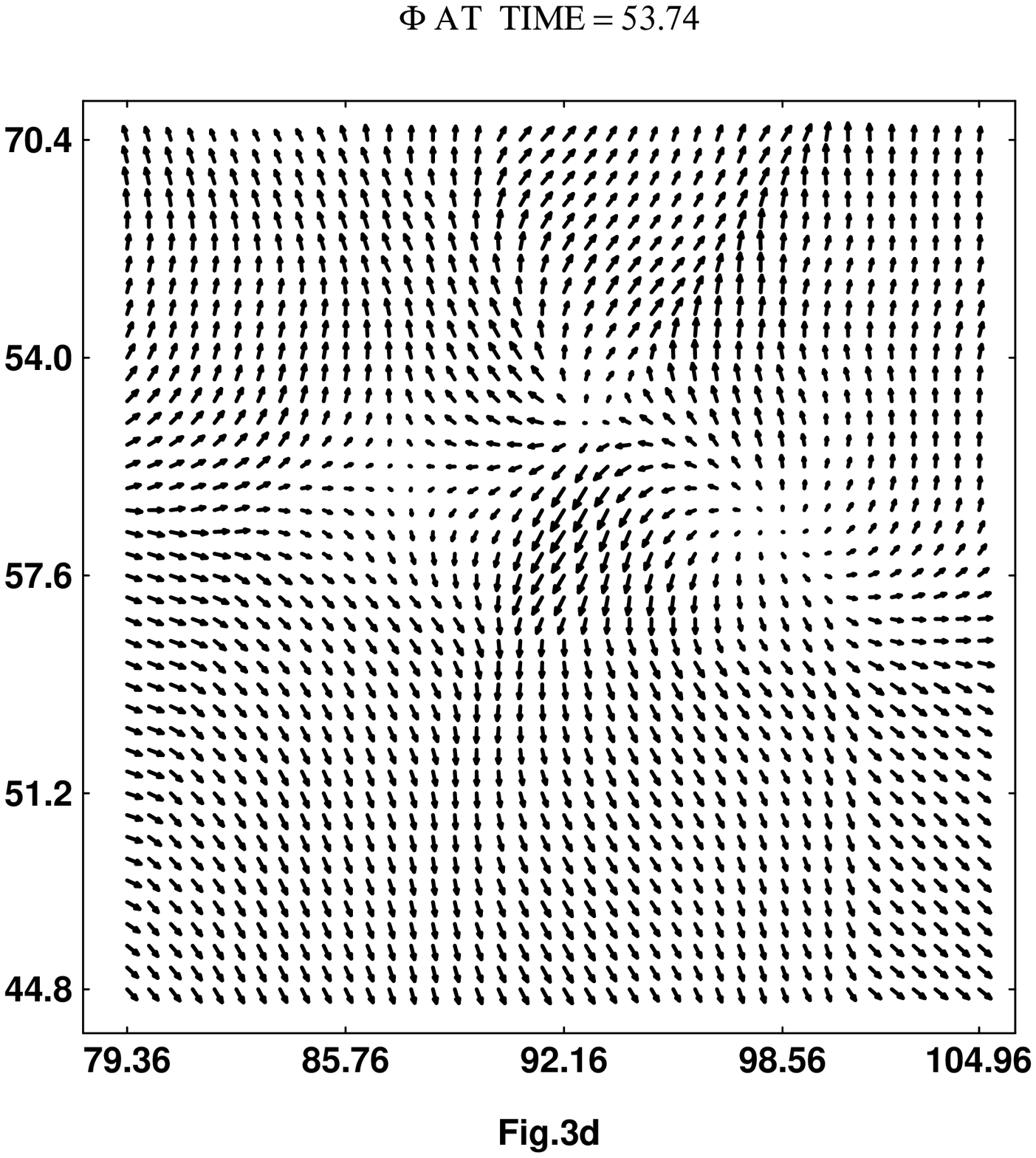}}
\vskip -0.5in
\end{center}
\caption{(a) A well separated vortex-antivortex pair. (b) The vortex 
and the antivortex are moving towards each other, while the pair, as
a whole, moves upwards. (c) Pair has annihilated by this time, as shown 
by the absence of any winding in this plot. (d) Vortex-antivortex pair
has been re-created. Note that the positions of the vortex and the 
antivortex have been exchanged.}
\label{Fig.3} 
\end{figure}

 Another interesting result we find is for the cases of certain
two bubble collisions which lead to vortex-antivortex pairs which 
are not well separated. However, due to presence of other bubbles,
density waves arising from other bubble collisions propagate
through the region where a given pair is getting formed.
This density wave then strongly affects the $\phi$ oscillations 
in that region and effectively separates the vortex and the
antivortex. Thus, due to the effects of density wave, a pair 
which was strongly overlapping and was going to annihilate soon, 
ends up in well separated and well formed vortex and antivortex.
This again shows that this new mechanism of pair creation strongly
depends on the dynamical details of the transition. A density wave
which as such can not contribute to defect production, leads to
enhancement in the production of well separated defects. Similar
effects were also found in ref. \onlinecite{dgl3}. 

\vskip .3in
\centerline{\bf IV. DEFECT PRODUCTION WITH}
\centerline{\bf  BIASED $\theta$ DISTRIBUTION}
\vskip .1in

  In this section we carry out a simulation  as in Section III, but
now with inclusion of a suppression factor in the nucleation rate for
bubbles with a given $\theta$ inside, resulting from the explicit
symmetry breaking. As we will see, Kibble mechanism vortices get 
strongly suppressed. To get an idea of this suppression, we carry
out a larger simulation with nucleation of 63 bubbles, while average
inter-bubble separation is the same as in Section III. This enables 
us to make direct comparison of average defect production per bubble
for the two cases (as defect production via the new mechanism depends
on average separation of bubble nucleation sites). 

 We estimate the suppression factor for bubble nucleation, with a
given $\theta$ inside, in the following manner. For this we assume 
that bubble nucleation is happening at some finite temperature $T$. 
(Even though the bubble profiles calculated in Section II correspond
to $T = 0$ case, we continue to use the same profiles. The reason for
this is that we want to change only one factor here from the simulation
of Section III, that is the $\theta$ suppression factor. In any case,
these are {\it thick wall} bubbles for which the profiles for $T = 0$ and 
$T \ne 0$ do not differ much. Also, actual profile of bubbles
which is relevant for us is at the time when bubbles start coalescing.
We are working with low nucleation rates, so bubbles expand by large
amount before coalescing. Thus initial difference in the profiles
becomes irrelevant.) This will lead to an additional factor 
$\Gamma_\theta$ in the nucleation rate of the bubbles which arises
due to the dependence of the effective potential on $\theta$. We have

\begin{equation}
\Gamma_\theta  = e^{-F(\theta)/T}
\end{equation}

\noindent where $F(\theta)$ is the contribution to the free energy
of the bubble arising from the explicit symmetry breaking term. For
$T$ we use the constraint that $T$ should be less than typical energy
required to create a vortex-antivortex pair, otherwise use of field
equations for the evolution will become suspect. We thus estimate the
energy of a pair of vortex-antivortex at various separations. For
this we use a code for minimization of energy as was used in 
ref.\onlinecite{emin}. To determine the energy of a vortex-antivortex
pair, at a given separation, one starts with a trial profile for
the pair. Field configuration is then fluctuated, while fixing the
centers of the vortex and the antivortex, and energy is minimized.
The configuration with the lowest value of energy is finally accepted 
as the correct profile of the vortex-antivortex pair and corresponding 
energy is taken as the energy of the pair. For details of the numerical 
technique, we refer the reader to ref.\onlinecite{emin}. 

 We find that at separation of about $m_H^{-1}$ (between the centers 
of the vortices) the energy of the pair, $E_{pair}$, is about 4.2 $m_H$, 
while at separation of 2$m_H^{-1}$ and 3$m_H^{-1}$, $E_{pair}$ 
increases to about 5.8 $m_H$ and 6.0 $m_H$ respectively. For the 
consistency of using field equations for the evolution, we require 
$T < E_{pair}$. To get $F(\theta)$, we integrate the explicit symmetry 
breaking term in Eqn.(1) for the profile of the bubble (determined 
from solving Eqn.(2)). We find $F(\theta) \simeq - 3.6 ~m_H cos\theta$. 
We choose a sample value of $T$ (satisfying the constraint that $T < 
E_{pair}$) as $T \simeq 1.8 ~m_H$. Resulting value of $F(\theta)/T$ is 
then simply equal to $- 2.0~ cos\theta$.
With this, we use the following (suitably normalized) expression for 
$\Gamma_\theta$ for determining bias in $\theta$ distribution in bubble 
nucleations.

\begin{equation}
\Gamma_\theta  = e^{2(cos\theta -1)}
\end{equation}

 With this choice of normalization, $\Gamma_\theta$ is 1 for 
$\theta = 0$ and is lowest, equal to $e^{-4}$, for $\theta = \pi$. 
 
  We then nucleate bubbles, with inclusion of this factor for 
$\theta$ suppression. To have reasonable statistics (due
to strong suppression of Kibble vortices in this case) we carry
out simulation over a larger physical region with size 
equal to 576 $\times$ 576. Total number of bubbles nucleated was
63 which lead to about the same average bubble separation as in
Section III. Just as in Section III, bubble locations are
determined randomly, but now $\theta$ inside bubbles is chosen 
with the weight factor $\Gamma_\theta$ as give in Eqn.(4).

  Resulting $\theta$ distribution of the bubbles is shown in 
Fig.4. Solid dots show the frequencies $P(\theta)$ of different 
values of $\theta$ obtained in the simulation, normalized so that the 
maximum value of $P(\theta)$ is 1.0. Solid curve is the plot of 
$\Gamma_\theta$.  Most important thing to
note in Fig.4 is an almost complete absence of $\theta$ points 
in the range $3\pi/2 > \theta > \pi/2$. In fact, there were
three bubbles with $\theta$ being slightly less than $3\pi/2$,
or slightly more than $\pi/2$. However, we verified that
very quickly, during the evolution and much before any other
bubble collides with these, $\theta$ in these bubbles rotates
(due to explicit symmetry breaking) and also falls in the range 
$3\pi/2 < \theta < \pi/2$. It is then immediately obvious that 
there is no possibility of any Kibble mechanism vortex forming
by coalescence of these bubbles as no winding can be generated by
following the geodesic rule. We have also verified it explicitly, 
by considering detailed field configurations of the vortices.

\begin{figure}[h]
\begin{center}
\vskip -0.5 in
\leavevmode
\epsfysize=10.0truecm \vbox{\epsfbox{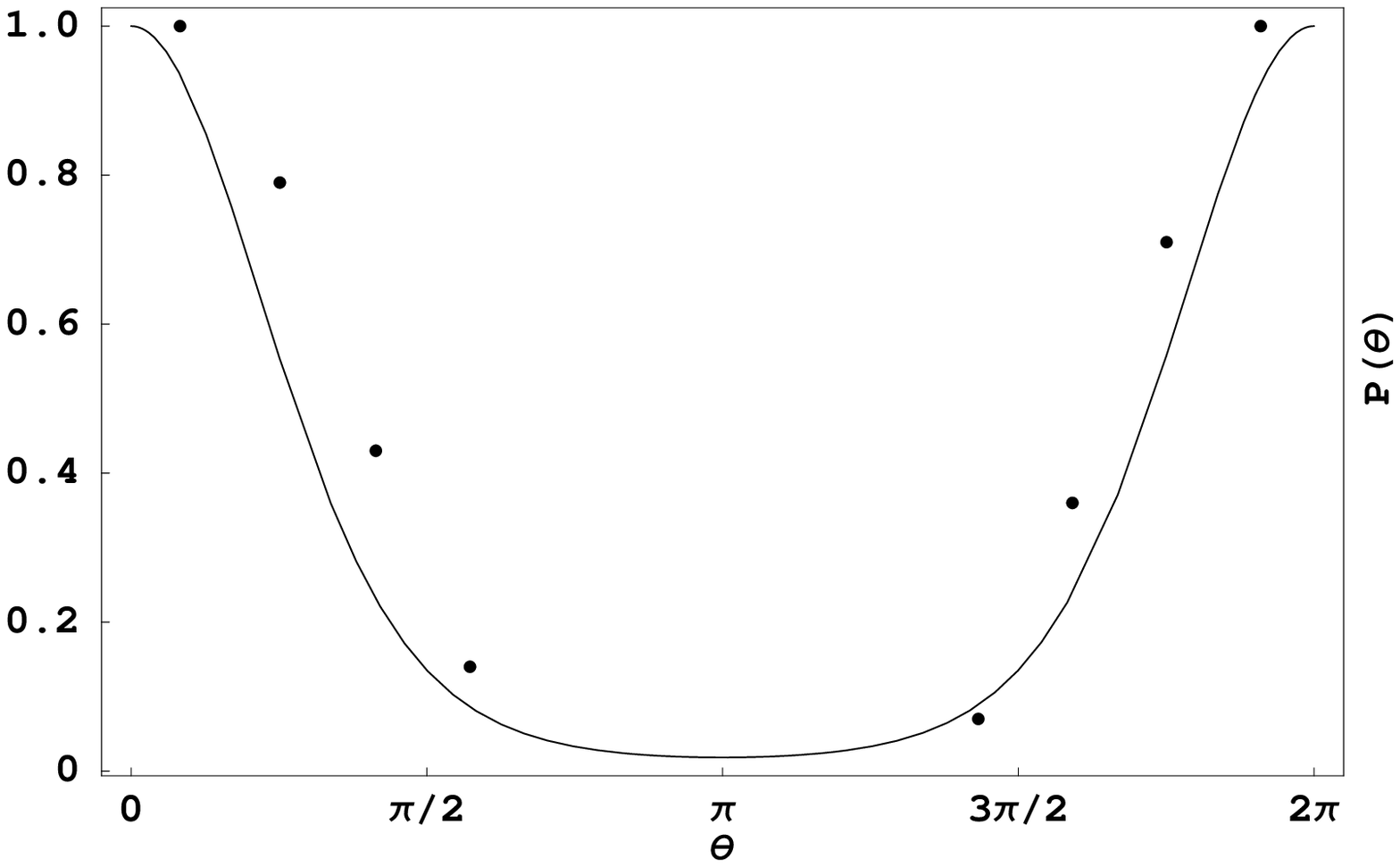}}
\vskip -1.5in
\end{center}
\caption{$\theta$ distribution inside bubbles. Dots show the
normalized frequency $P(\theta)$ of $\theta$ inside different 
bubbles nucleated in the simulation. Solid curve is the plot of 
$\Gamma_\theta$.}
\label{Fig.4} 
\end{figure}

  We get a total of 52 vortex-antivortex pairs (i.e. 104 vortices
and antivortices), all due to the new mechanism. This gives the
average defect production per bubble to be about 1.65 which is
essentially the same as the average number of defects per bubble 
via the new mechanism, as found in the simulation in 
Section III. Thus biased $\theta$ 
distribution does not seem to adversely affect defect production via 
this new mechanism. This results is important as it shows that in 
certain situations, with explicit symmetry breaking, this mechanism 
may not just dominate, it may be the only mechanism active for defect 
production. Effect of biased $\theta$ distribution on this mechanism 
can be understood as follows. Suppression of larger values of $\theta$
leads to smaller potential energy being stored in the collision
regions, and this should adversely affect wall oscillation and
hence defect production. However, now average $\theta$ difference
between coalescing bubbles will be smaller. As mentioned earlier,
this helps in increasing wall oscillations, as less energy is spent
in overcoming $\theta$ gradients. This should lead to increase in
defect production. Final defect production will be a combination of
both these factors.

 In next sections, we will study effects of changing nucleation
rate, magnitude of explicit symmetry breaking, and presence of
damping, on defect production via this new mechanism. As $\theta$ 
suppression does not seem to affect this mechanism much, we will
not include its effect in these sections. This also helps in
making conservative comparison of this mechanism with the Kibble
mechanism, when these parameters are varied, as in absence of
any bias in $\theta$ distribution Kibble vortices are also 
unsuppressed.

\vskip .3in
\centerline{\bf V. DEPENDENCE OF DEFECT PRODUCTION}
\centerline{\bf ON NUCLEATION RATE AND $\kappa$}
\vskip .1in

   The results discussed in last two sections show that there is 
a strong enhancement in the number of defects produced in a phase 
transition due to this new mechanism. In this section we will
study the dependence of this enhancement in defect production
on the magnitude of symmetry breaking coefficient $\kappa$ 
and on the nucleation rate.  We start by discussing the effects
of changing the nucleation rate, which in effect changes the
average separation between bubbles.

  We use $\kappa=0.015$ as in Section III, and keep it
fixed while we carry out simulations with different values of the
nucleation probability. As we are now using uniform probability
for $\theta$ distribution inside bubbles, we will compare the
results of this section with those in Section III.
Bubble nucleation probability used for the
case discussed in Section III lead to seven bubbles nucleated
in the lattice. We have studied two other cases by changing 
the nucleation probability. In one case the net number of bubbles
nucleated was 12 and in the other case number of bubbles was 20.

 For the case of 12 bubbles we found a total of 18 vortices.
Out of these, 3 pairs of vortex-antivortex formed a cluster. For 
this simulation the total defect density per bubble comes out to be  
1.5 which is less then the case of seven bubbles discussed in the 
previous section. This is consistent with the physical picture of
this mechanism we discussed earlier where we mentioned that
energetic bubble collisions will lead to larger oscillations of $\phi$
in the coalesced portions of bubble walls
and consequently to larger number of vortex-antivortex pairs getting
produced. Twelve bubble case leads to smaller value of average
separation between bubbles compared to the seven bubble case
and hence less energy for $\phi$ oscillations.

  As for the case of seven bubbles of Section III, here also
we find several additional pairs of vortex-antivortex which 
correspond to strongly overlapping vortex-antivortex 
configurations. These did not separate and lasted only for
a short time. There were three such pairs in the present
case of twelve bubbles but we did not count them in the 
total number of vortices. 

For the second case we take the value of nucleation probability 
so that there are twenty bubbles nucleated in the whole lattice.
Since the average separation between the bubbles is smallest in 
this case compared to the previous two cases, the field
oscillations flipping $\Phi$ will be weakest in this case.
Thus we expect that the pair production mechanism will not
lead to many defects in this case. This is precisely what we 
find as the total number of vortices and antivortices in this case
comes out to be 14. This gives the number density of all defects 
per bubble to be equal to 0.7. Again, apart from these 14 defects,
we found six vortex-antivortex pairs which are strongly 
overlapping and therefore are not counted in determining the net 
number of defects. These results are similar to those found in
ref. \onlinecite{dgl3} where larger nucleation rate lead to a 
smaller number of well separated pairs being produced.

To study the effect of $\kappa$ on the number density of defects
we took the initial field configuration to be that corresponding
to the case of twelve bubbles mentioned above. We then carry
out simulations for various values of $\kappa$, each time starting
with exactly the same initial field configuration.  It is easy to 
see that increasing $\kappa$ will lead to faster rolling of $\theta$ 
inside the bubbles. From this one may have the impression that
larger values of $\kappa$ may lead to smaller number of defects 
as $\theta$ will be close to zero in all the bubbles.  As mentioned 
in ref. \onlinecite{dgl1}, this indeed does happen for values of 
$\kappa$ which are so large that $\theta$ in the bubbles settles 
down to zero before the $\phi$ oscillations can take place.  For our 
choices of parameters in Eqn.(3), this happens for values of $\kappa$ 
larger than 0.03, see ref. \onlinecite{dgl1}.

 However, as long as $\kappa$ is smaller than this value, an increase 
in its value leads to increase in the defect production. It is easy to 
understand why this happens.  Increasing $\kappa$ leads to increase in 
the potential energy, as well as the gradient energy  
of the field configuration in the coalesced region. For example,
the region where $\theta$ takes value $\pi$ has much larger
energy with larger value of $\kappa$. There is, thus, more
energy available for wall to decay in vortex-antivortex pairs
leading to larger number of defects.

  The case of twelve bubbles which we described above corresponded
to the value of $\kappa$ equal to 0.015. We repeated that simulation
(with exactly same initial field configuration) for two other
values of $\kappa$, one with $\kappa$ equal to 0.01 and the other
with $\kappa$ equal to 0.02. For the case with $\kappa$ equal to
0.01 we got a total of 12 vortices and antivortices giving
the average number of defects per bubble to be 1.0. This should
be compared with the defect density per bubble = 1.5 for
$\kappa = 0.015$ case. From the above discussion, this decrease 
in defect production for smaller value of $\kappa$ is expected.
For $\kappa = 0.02$ we found a total of 22 vortices and
antivortices giving the density of defects per bubble to be
equal to 1.8, clearly a significant increase. In this series of 
simulations for different values of $\kappa$ we found that some of 
the two bubble collisions lead to vortex-antivortex pairs only for 
$\kappa=0.02$. (Note that initial bubble configurations have been
chosen to be identical for all the three cases.) Which implies that 
$\phi$ oscillations were large enough to flip $\Phi$ only for this 
larger value of $\kappa$.

 There is one important feature we found which needs to be 
emphasized. In the case of $\kappa = 0.01$ (for twelve bubble 
case), we found that out of the total of 12 vortices, eight 
vortices can be attributed to be forming from the Kibble Mechanism 
and four from the pair production process. However, when the simulation 
was repeated for $\kappa=0.0$ (again starting with exactly same initial 
field configuration) then only four vortices formed via the Kibble
mechanism.  The reason that the vortices corresponding to the Kibble 
mechanism are fewer in this case is that these corresponded to collisions
of more than three bubbles. Altogether these bubbles had zero net winding 
but taken in groups of three bubbles at a time, they had winding and 
anti-winding (using the geodesic rule). In the case of $\kappa = 0.0$
this vortex and antivortex  immediately annihilate each other (or
in some sense their windings annihilate each other even before
they could form). However, for non-zero $\kappa$, this vortex and
antivortex get separated  in order to minimize the energy as
$\theta$ in between the vortex and antivortex takes the value
zero which is the absolute minimum of the effective potential.
Thus explicit symmetry breaking, in addition to producing defects 
via wall decaying into pairs, also can enhance the defect production 
via the Kibble mechanism in certain situations. 

To summarize the results of this section, we found that increasing 
the nucleation rate decreases the number density of vortices per 
bubble which is due to decrease in average separation between 
the bubbles. Smaller separation between bubbles leads to a
suppression in $\phi$ oscillations and consequently decreases the
defect production. Increase in the magnitude of the explicit
symmetry breaking $\kappa$,  increases the number density of vortices 
per bubble. This is attributed to  enhancement in $\phi$ oscillation 
coming from the increase in potential energy and gradient energy of 
wall configuration in the coalesced region due to a larger $\kappa$.

\vskip .3in
\centerline{\bf V. EFFECT OF DAMPING ON THE}
\centerline{\bf  NEW MECHANISM}
\vskip .1in

 So far our discussion (as well as the discussion in ref. 
\onlinecite{dgl1}) was for the case with no dissipation present.
Since the creation of vortex-antivortex pairs via this 
mechanism crucially depends on $\phi$ oscillations (and 
on $\theta$ gradient which comes from $\theta$ oscillations),
it is natural to expect that the presence of damping can
crucially affect the effectiveness of this mechanism. It is
thus important to understand the precise manner in which
damping affects this mechanism.  After all, damping is naturally 
present in most cases of interest, such as the early Universe, 
liquid crystal systems and even quark-gluon plasma at finite 
temperature (for the case of Skyrmions). 

In this section we will present results of simulations of two 
bubble collisions in presence of dissipation and study how the 
defect production is affected. What we expect to find is that the
presence of damping should suppress the defect production
via this mechanism. This is because damping will lead to suppression
in the amplitudes of successive oscillations of $\phi$ in the wall
so that it will become difficult for $\Phi$ to flip. Consequently,
production of vortex-antivortex pairs will be suppressed. This is what 
we find in our simulations, which we describe in the following.

        We have studied the effect of damping on the creation of 
vortices in two bubble collision by introducing a damping term 
$\eta \dot{\phi}$ in the equation of motions for evolving 
the field configuration. Lattice size is taken to be 160 x 128, with
the same lattice spacing as before.
We choose the initial field configuration
of the two bubbles to be the one which leads to formation of three 
vortex-antivortex pairs when damping is absent. Value of $\kappa$ is 
taken to be equal to 0.015. We then  repeat the simulation for this two 
bubble collision starting with exactly the same initial configuration, 
but now in the presence of damping. We study the effect of magnitude of 
damping by changing the damping coefficient $\eta$. In the presence of 
damping, bubble wall velocity is smaller so the bubble collision is less 
energetic to begin with. Further, as the oscillating $\phi$ in the 
coalesced region looses energy also due to damping, successive 
oscillations are strongly suppressed. This leads to suppression in the 
number of pairs getting formed. 

  For small values of $\eta$ (in the range 0.0 to 0.25) we still get 
three vortex-antivortex pairs. But for the values of $\eta$ larger than 
0.25, $\phi$ oscillations are strongly suppressed and only one pair is 
formed. For $\eta = 1.0$, even the one pair formed corresponds to a 
vortex and an antivortex which are very close to the wall of the 
coalesced bubbles and are not well formed. $\phi$ oscillations are 
very strongly suppressed in this case and $\theta$ almost settles 
down to zero without any oscillation about $\theta=0.0$. Figs.5a and 
5b show surface plots of $-\phi$ of the coalesced region of the two 
bubbles, at the same stage, $t = 84.85$, for the two cases, $\eta = 
0.1$ and $\eta = 0.75$ respectively.  Starting $\theta$ in both 
the bubbles was such that $\theta$ in between the bubbles 
interpolates through $\theta = \pi$. As we mentioned above, the 
initial field configurations of the two bubbles were identical for 
both of these cases. We see that there are two, well formed, 
vortex-antivortex pairs present in Fig.5a, for $\eta = 0.1$ case.  
In contrast, at the same stage, there is only one pair present in 
Fig.5b for $\eta = 0.75$ case. For $\eta = 0.1$ case, one more pair 
forms later, making total number of pairs formed equal to 3 for this 
case. However, by that time, the only pair formed  for the $\eta = 
0.75$ case escapes out of the walls of the coalesced bubble (due to 
$\theta$ being zero between the pair, and $\theta = \pi$ in directions 
towards the walls), hence we do not show plots at that stage. 

\begin{figure}[h]
\begin{center}
\vskip -1.0 in
\leavevmode
\epsfysize=10truecm \vbox{\epsfbox{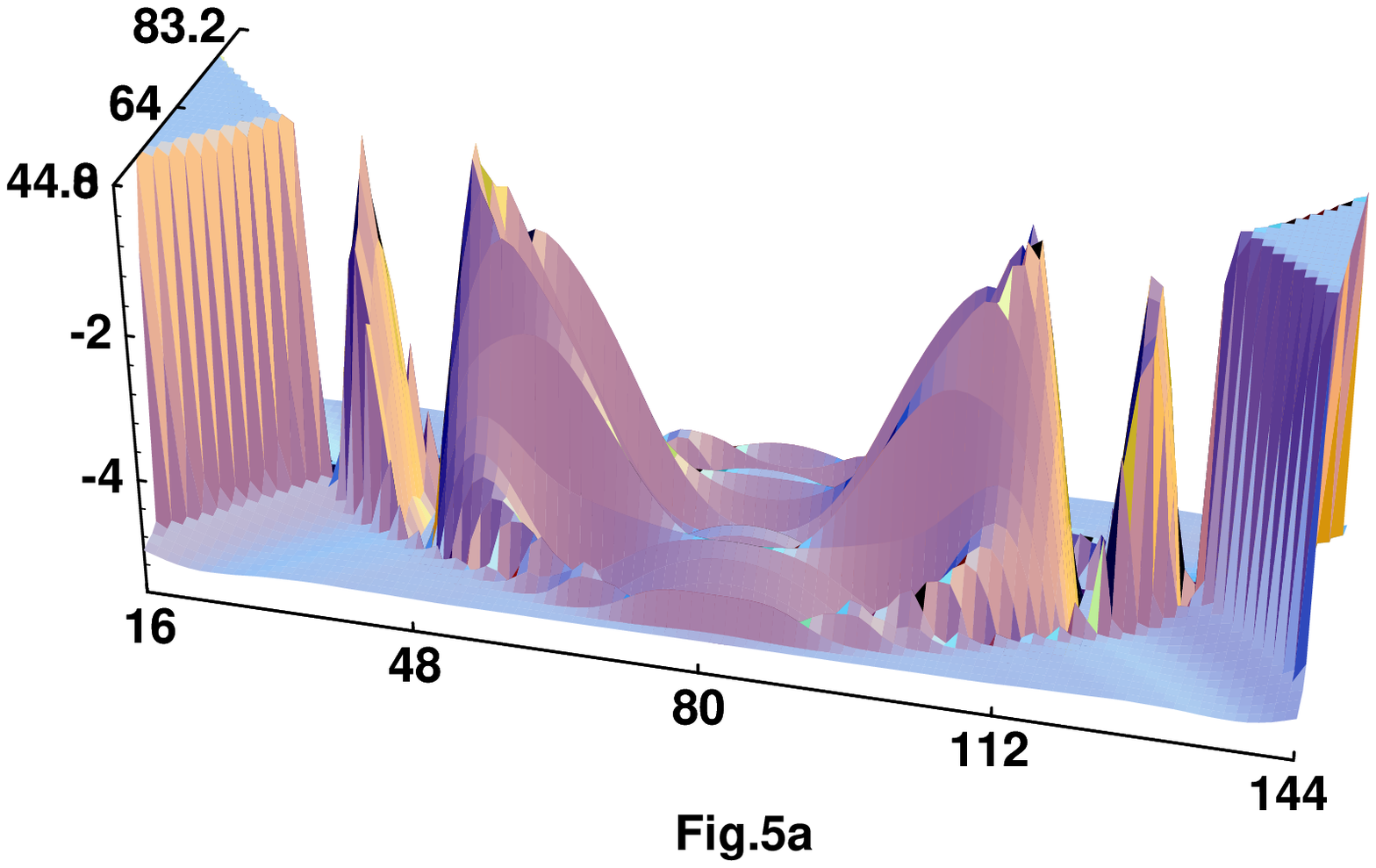}}
\vskip 0.5in
\end{center}
\end{figure}

\begin{figure}[h]
\begin{center}
\vskip -3 in
\leavevmode
\epsfysize=10truecm \vbox{\epsfbox{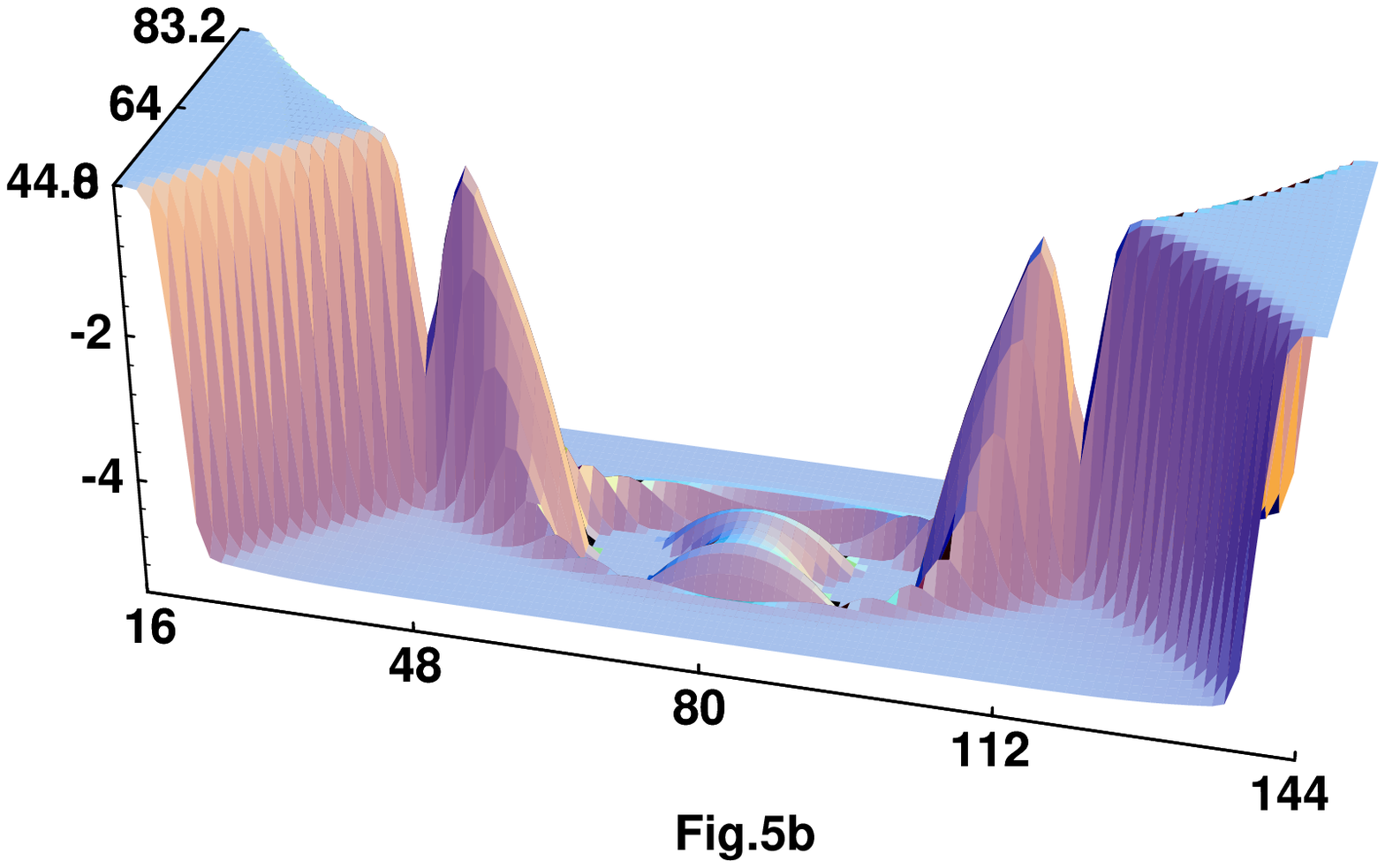}}
\vskip -1.0in
\end{center}
\caption{(a) Surface plot of $-\phi$ for $\eta = 0.1$ of the coalesced
region of two bubbles, showing profiles of two vortex-antivortex pairs. 
One more pair forms later in this case. (b) Surface plot of $-\phi$ for 
$\eta = 0.75$ showing the profile of the single pair formed in this 
case. Later, this vortex and the antivortex escape out of the coalesced 
bubble.  Both these plots are shown at the same time stage, $t = 84.85$, 
starting with identical initial field configurations of the two bubbles.}
\label{Fig.5} 
\end{figure}

 We emphasize here that whatever be the value of the damping 
coefficient, there is always  at least one  vortex-antivortex pair 
formed with favorable initial configuration of two bubbles (i.e. 
$\theta$ distribution) via the new mechanism. This happens because 
the potential energy 
contained in the coalesced region will make the flipping of $\Phi$
(and hence the pair production) favorable compared to rolling down 
of the two ends of the arc in Fig.1 as that leads to constantly 
increasing gradient energy. We also mention that irrespective of
damping (or even the order of phase transition) one pair should always 
form, for appropriate values of $\theta$ in the two bubbles, via the 
mechanism discussed in ref. \onlinecite{kpst} as $\phi$ oscillations
play no role there. However, we have checked that rotation of $\theta$ in
the walls is so slow (due to the presence of $\phi^2$ in the explicit
symmetry breaking term) that one will expect the pair formation via the 
mechanism of ref. \onlinecite{kpst} only after a very long time. On the 
other hand, pair formation due to $\phi$ flipping happens in a much 
shorter time scale, that of a single $\phi$  oscillation in the wall.

 Another important point which should be mentioned here is that, in the 
presence of strong damping, no pair may get created if $\kappa$ is 
too small. In the presence of strong damping one can neglect the kinetic 
energy of the bubble wall. So what remains is the contribution from the 
explicit symmetry breaking term, which must lead to the flipping of 
$\Phi$ by forcing it to go through zero. However, below a certain 
critical value of $\kappa$ it may not happen, as the $\theta$ gradient 
developed in the coalesced portion of the bubbles may form a stable 
domain wall. This will be like the axionic domain wall \cite{axn}, 
which can decay only via quantum tunneling (say, at zero temperature).

 This is an appropriate point to discuss an important issue about 
the applicability of the new mechanism for different types of defects. 
There is one class of defects where the applicability of this mechanism 
may hold only in some special situations, and not generically. These 
correspond to vacuum manifolds for which the opposite orientations of 
the order parameter field are identified. Nematic liquid crystals 
happens to belong to this category, with the vacuum manifold being 
$RP^2$. In such cases, flipping of the order parameter field does not 
change its configuration. Thus the argument given above for the pair 
production can not be directly applied here. However, it is possible  
to argue that in the presence of explicit symmetry breaking 
this mechanism should still be applicable, especially 
when the system is dissipative. Consider, for example, an 
order parameter configuration varying around a point P in the 
physical space smoothly such that the value of the order parameter at
P is energetically most unfavorable (due to explicit symmetry breaking).
For the case described by Eqn.(1) it means $\theta$ having value $\pi$
at point P, and varying from a value less than $\pi$ to a value larger
than $\pi$ as we cross P in the physical space. 

 For nematic liquid crystals, say, in the presence of external electric 
field along the x axis, it means that the director (order parameter) 
will lie in the y-z plane at the point P, and will deviate from the 
y-z plane in opposite directions as one passes through P in the 
physical space.  As the order parameter field rolls down to the direction 
of the true vacuum, a region of large gradient energy will 
arise near P. As we explained above, for sufficiently small explicit
symmetry breaking, the gradient energy may not become too large, and
may result in the formation of a domain wall in that region. However, 
for larger values of explicit symmetry breaking (e.g., stronger 
electric field for the liquid crystal case) , the gradient energy will
keep rising, eventually forcing the field to go to zero magnitude in the 
region near P. The only way to decrease the energy of this configuration 
is by creating a defect-antidefect pair. For extremely damped situations
it may happen gradually as the director reaches zero magnitude at P and 
then slowly grows in magnitude but now with an orientation along the x 
axis, thereby resulting in the creation of vortex-antivortex pair 
(for 2-dimensions, and a loop of string for 3-dimensions). For smaller 
damping, or for that matter, in the absence of damping, the order 
parameter field may keep oscillating at P through zero,  without 
changing its configuration in passing through zero. During the passage 
of the field through zero, any small fluctuation can re-orient the 
director along the x axis, resulting in the pair production.  Or, the 
field may eventually settle at zero (either due to the damping term, or 
just due to loss of energy in successive oscillations in other 
excitations of the system), and then {\it roll down} to a direction 
along the x axis. This will then again result in the pair production 
via this mechanism. In the absence of any explicit symmetry breaking, 
this mechanism does not seem to be applicable for these types of 
vacuum manifolds. 
  
\vskip .3in
\centerline {\bf VI. CONCLUSIONS}
\vskip .1in
 
   We conclude by emphasizing the essential aspects of our results.  We 
have carried out numerical simulations of a first order phase transition 
by random nucleation of bubbles, in the presence of small explicit 
symmetry breaking, and have studied the production of vortices and 
antivortices.  We estimate the net number density (number of defects
per bubble) of vortices produced,
which includes vortices formed due to the Kibble mechanism
as well as those produced via the pair production mechanism.
We also study the dependence of this defect number density on
parameters such as the magnitude of the explicit symmetry breaking
term as well as on the nucleation rate. Defect production increases
with larger magnitude of explicit symmetry breaking due to larger
potential energy in the coalesced region. Nucleation rate affects
defect density due to the fact that a larger nucleation rate 
implies smaller average bubble separation, which in turn leads
to less kinetic energy for the bubble walls before bubbles
collide. Oscillations of $\phi$ are less prominent for less
energetic walls leading to smaller number of defects for larger
nucleation rates. We have also studied the effects of presence
of damping on this pair production mechanism by studying pair
production in two bubble collisions. As expected, we find the damping
suppresses oscillations of $\phi$ and hence leads to smaller number of 
defects, though one pair is always produced (for suitable values of
$\theta$ distribution in the bubbles, and for $\kappa$ not too small).

  In all these cases we find that the number of defects produced
due to effects of explicit symmetry breaking term (either via direct pair 
production mechanism, or by separating vortex-antivortex pairs produced
in  collisions of more than three bubbles via the Kibble mechanism) is 
much larger than what one would expect from the Kibble mechanism. 
This relative enhancement is much larger when we include the effect of
bias in $\theta$ distribution in bubbles, resulting from explicit symmetry 
breaking. In this case we find, in a simulation with 63 bubbles, that not 
even a single defect is produced via the Kibble mechanism, while 104 
vortices and antivortices are produced via the new mechanism. As the 
distribution of these defects is of very different nature than the one 
expected from the Kibble mechanism, one may expect qualitatively new 
features for systems with explicit symmetry breaking. In our 2+1 
dimensional study we find a sequence of vortex-antivortex pairs being 
produced. Simple arguments show that for 3+1 dimensions one will get  
a system of concentric string loops, (see ref. \onlinecite{dgl1} for 
example).  This may then suppress formation of large strings compared 
to small loops. It will be of great interest if such a prediction can be 
experimentally verified in some physical system. As mentioned in 
ref. \onlinecite{dgl1}, formation of strings in nematic liquid crystals
in the presence of external electric or magnetic fields may be ideal
for checking this mechanism. (Though in that case, presence of damping 
may suppress string formation via this new mechanism.) 

\acknowledgements

 We are very grateful to Ed Copeland and Paul Saffin for very useful
comments, especially for pointing out to us the non-trivial issue of
correct bounce solution for the case with explicit symmetry breaking. 
We also thank an anonymous referee for useful comments about
constraints on temperature for bubble nucleation in presence of
explicit symmetry breaking, and about $\theta$ suppression
factor in bubble nucleation.

\end{multicols}

\begin{thebibliography}{99}

\bibitem{shlrd} For a review see, A. Vilenkin and E.P.S. Shellard,
``Cosmic strings and other topological defects", (Cambridge
University Press, Cambridge, 1994).

\bibitem{thrm} F.A. Bais and S. Rudaz, Nucl. Phys.  {\bf B170}, 507 
(1980); F. Liu, M. Mondello, and N. Goldenfeld, Phys. Rev. Lett. 
{\bf 66}, 3071 (1991).  
\bibitem{kbl} T.W.B. Kibble, J. Phys. {\bf A9}, 1387 (1976).

\bibitem{zurk} For a review, see, W.H. Zurek, Phys. Rep. {\bf 276}, 
177 (1996). 

\bibitem{lcrs} S. Digal, R. Ray, and A.M. Srivastava, 
hep-ph/9805502.

\bibitem{cs1} E.J. Copeland and P.M. Saffin, Phys. Rev. {\bf D54},
6088 (1996). 

\bibitem{brl} J. Borrill, T.W.B. Kibble, T. Vachaspati, and A. Vilenkin,
Phys. Rev. {\bf D52}, 1934 (1995).

\bibitem{dgl1} S. Digal and A.M. Srivastava, Phys. Rev. Lett. {\bf 76}, 
583 (1996).

\bibitem{dgl2} S. Digal, S. Sengupta, and A.M. Srivastava, Phys. Rev. 
{\bf D55}, 3824 (1997).

\bibitem{dgl3} S. Digal, S. Sengupta, and A.M. Srivastava,  Phys. Rev. 
{\bf D56}, 2035 (1997).

\bibitem{cs2} E.J. Copeland and P.M. Saffin, Phys. Rev. {\bf D56},
1215 (1997). 

\bibitem{volov} G.E. Volovik, Czech. J. Phys. {\bf 46}, 3048 (1996) 
Suppl. S6; T.D.C. Bevan, A.J. Manninen, J.B. Cook, J.R. Hook, H.E. 
Hall, T. Vachaspati, and G.E. Volovik, Nature, {\bf 386}, 689 (1997).  

\bibitem{kpst} J.I. Kapusta and A.M. Srivastava, Phys. Rev. 
{\bf D52}, 2977 (1995).

\bibitem{axn} For a detailed discussion see, E.W. Kolb and
M.S. Turner, ``The early Universe", (Addison-Wesley, Redwood
City, California, 1990). 

\bibitem{lc} P. G. de Gennes, ``The physics of liquid crystals",
(Clarendon, Oxford, 1974).

\bibitem{ajit} A.M. Srivastava, Phys. Rev. {\bf D45}, R3304 (1992);
{\bf D46}, 1353 (1992); S. Chakravarty and A.M. Srivastava,
Nucl. Phys. {\bf B406}, 795 (1993).

\bibitem{bbl} M.B. Voloshin, I.Yu. Kobzarev and L.B. Okun,
Yad. Fiz. {\bf 20}, 1229 (1974) [Sov. J. Nucl. Phys. {\bf 20},
644 (1975)]; S. Coleman, Phys. Rev. {\bf D15}, 2929 (1977).

\bibitem{bbl2} E.J. Copeland and P.M. Saffin, private communication.

\bibitem{carl} C. Rosenzweig and A.M. Srivastava, Phys.
Rev. {\bf D 43}, 4029 (1991).

\bibitem{mmr} K.J.M. Moriarty, E. Myers and C. Rebbi,
Phys. Lett. {\bf B207}, 411 (1988).

\bibitem{emin} A.M. Srivastava, Phys. Rev. {\bf D 47}, 1324 (1993).

\end{thebibliography}
\end{document}